\documentclass[twocolumn,Astrosymb]{aastex61}

\usepackage{natbib}

\submitjournal{PASP}

\shorttitle{NIRISS Observations of TESS planets}
\shortauthors{Louie et al.}

\begin{document}

\title{Simulated JWST/NIRISS Transit Spectroscopy of Anticipated TESS Planets\\Compared to Select Discoveries from Space-Based and Ground-Based Surveys}

\correspondingauthor{Dana R. Louie}
\email{danalouie@astro.umd.edu}


\author{Dana R. Louie}
\affiliation{Department of Astronomy, University of Maryland, College Park, MD 20742, USA}

\author{Drake Deming}
\affiliation{Department of Astronomy, University of Maryland, College Park, MD 20742, USA}
\affiliation{TESS Atmospheric Characterization Working Group}

\author{Loic Albert}
\affiliation{Institut de recherche sur les exoplanetes (iREx), Universite de Montreal, Montreal, Quebec, Canada}

\author{L. G. Bouma}
\affiliation{Department of Astrophysical Sciences, Princeton University, Princeton, NJ 08540, USA}

\author{Jacob Bean}
\affiliation{Department of Astronomy and Astrophysics, University of Chicago, IL 60637, USA}
\affiliation{TESS Atmospheric Characterization Working Group}

\author{Mercedes Lopez-Morales}
\affiliation{Harvard-Smithsonian Center for Astrophysics, Cambridge, MA 02138, USA }
\affiliation{TESS Atmospheric Characterization Working Group}



\begin{abstract}

The Transiting Exoplanet Survey Satellite (TESS) will embark in 2018 on a 2-year wide-field survey mission, discovering over a thousand terrestrial, super-Earth and sub-Neptune-sized exoplanets ($R_{\rm pl} \leq 4R_{\oplus}$) potentially suitable for follow-up observations using the James Webb Space Telescope (JWST).  This work aims to understand the suitability of anticipated TESS planet discoveries for atmospheric characterization by JWST's Near InfraRed Imager and Slitless Spectrograph (NIRISS) by employing a simulation tool to estimate the signal-to-noise (S/N) achievable in transmission spectroscopy. We applied this tool to Monte Carlo predictions of the TESS expected planet yield and then compared the S/N for anticipated TESS discoveries to our estimates of S/N for 18 known exoplanets.  We analyzed the sensitivity of our results to planetary composition, cloud cover, and presence of an observational noise floor.  We find that several hundred anticipated TESS discoveries with radii $1.5R_{\oplus} < R_{\rm pl} \leq 2.5R_{\oplus}$ will produce S/N higher than currently known exoplanets in this radius regime, such as K2-3b or K2-3c.  In the terrestrial planet regime, we find that only a few anticipated TESS discoveries will result in higher S/N than currently known exoplanets, such as the TRAPPIST-1 planets, GJ1132b, and LHS1140b.  However, we emphasize that this outcome is based upon Kepler-derived occurrence rates, and that co-planar compact multi-planet systems (e.g., TRAPPIST-1) may be under-represented in the predicted TESS planet yield.  Finally, we apply our calculations to estimate the required magnitude of a JWST follow-up program devoted to mapping the transition region between hydrogen-dominated and high molecular weight atmospheres.  We find that a modest observing program of between 60 to 100 hours of charged JWST time can define the nature of that transition (e.g., step function versus a power law).  

\end{abstract}

\keywords{eclipses --- infrared: planetary systems --- instrumentation: spectrographs  --- planets and satellites: atmospheres --- space vehicles: instruments --- techniques: spectroscopic}



\section{Introduction} \label{sec:intro}

The approaching launch of the James Webb Space Telescope (JWST), coupled with the 2018 launch of the Transiting Exoplanet Survey Satellite (TESS), heralds a new era in exoplanet science, with TESS projected to detect over one thousand transiting exoplanets smaller than Neptune \citep{2014SPIE.9143E..20R}, and JWST offering unprecedented spectroscopic capabilities through which we can examine exoplanetary atmospheres \citep{2006SSRv..123..485G, 2014PASP..126.1134B, 2016ApJ...817...17G}.  

One goal of exploring exoplanet atmospheres is to identify biosignatures, thus firmly establishing whether life exists on planets orbiting other stars \citep{2010ARA&A..48..631S}.  Further, because water is necessary for life on Earth, we expect life will develop on planets able to maintain liquid water, and atmospheric water vapor can be used as a proxy for liquid surface water \citep{2013Sci...340..577S}.  Rather than focusing on true Earth analogs, \cite{2007arXiv0706.1047C} describe the advantages offered through atmospheric characterization of super-Earths orbiting in the habitable zone (HZ) of nearby M-dwarfs.  

\cite{2015ApJ...809...77S} used Monte Carlo simulations to predict that TESS will detect approximately 1,700 transiting planets orbiting pre-selected target stars during its 2-year wide-field survey of the northern and southern ecliptic hemispheres.  The simulations employ Kepler-derived planet occurrence rates, as well as photometric performance models for the TESS cameras.  Notably, \cite{2015ApJ...809...77S} found that about one-third of the TESS-discovered transiting exoplanets will have radii less than twice that of the Earth's, and three-quarters of these $R_{\rm pl} < 2R_\oplus$ planets will orbit M-dwarfs.  Slightly fewer than 10 percent of the $R_{\rm pl} < 2R_\oplus$ planets will orbit near or within their host star's habitable zone.  Exoplanets found by TESS will orbit stars 10-100 times brighter than those found during Kepler's primary mission, thus facilitating follow-up characterization of their atmospheres \citep{2014SPIE.9143E..20R}. 

Although TESS is poised to discover a multitude of M-dwarf-transiting sub-Neptune-sized and smaller exoplanets, other missions and ground-based projects have already discovered many favorable transiting planets.  NASA's Kepler spacecraft was repurposed to fulfill the K2 mission, which includes observation of transiting exoplanets orbiting bright, low-mass stars \citep{2014PASP..126..398H}.  Furthermore, ground-based surveys such as MEarth \citep{2015csss...18..767I, 2009IAUS..253...37I, 2008PASP..120..317N} and TRAPPIST \citep{2011Msngr.145....2J} have recently announced exciting exoplanet discoveries, such as that of GJ1132 \citep{2015Natur.527..204B}, the seven terrestrial-sized TRAPPIST-1 planets \citep{2016Natur.533..221G, 2017Natur.542..456G}, and the habitable zone super-Earth LHS-1140b \citep{2017Natur.544..333D}.  

The targeted planets for JWST atmospheric characterization studies must be chosen wisely to maximize the amount of scientific knowledge attained for a given amount of JWST observation time \citep{2017ApJ...835...96H, 2017AJ....153..151B}.  During its 2-year primary mission, TESS will discover exoplanets continually, and the first discoveries will not necessarily be those most conducive to follow-up atmospheric characterization.  

One motivation of this paper is to determine which TESS discoveries will produce the highest signal-to-noise (S/N) in transmission spectroscopy using the JWST Near-Infrared Imager and Slitless Spectrograph (NIRISS) operating in Single Object Slitless Spectroscopy (SOSS) mode.  To attain this goal, we simulate NIRISS instrument performance during observations of the anticipated TESS exoplanet discoveries, limiting our predictions to planets with $R_{\rm pl} < 4R_\oplus$.  We then compare our results for the TESS planets to similar predictions for selected exoplanets already discovered via space-based or ground-based surveys.  This comparison allows us to predict the highest priority TESS discoveries for immediate confirmation and follow-up observation. Furthermore, the TESS discoveries we examine lie within that regime where planets transition from rocky planets surrounded by high molecular weight atmospheres, to icy sub-Neptunes enveloped in hydrogen-dominated atmospheres \citep{2017arXiv170310375F}.  Thus, an additional motivation of our work is to predict the magnitude of the observational program required to map this transition region.

Numerous past studies have estimated JWST's capabilities during exoplanet atmosphere characterization \citep{2009PASP..121..952D, 2011IAUS..276..335C, 2011A&A...525A..83B, 2014PASP..126.1134B, 2015MNRAS.448.2546B, 2016ApJ...817...17G, 2016SPIE.9904E..3PB, 2017arXiv170103539C, 2017AJ....153..151B, 2017ApJ...835...96H, 2017A&A...600A..10M}, and NIRISS has emerged as the workhorse instrument for transit spectroscopy (e.g., \citealp{2016PASP..128i4401S}).  \cite{2016ApJ...817...17G} modeled archetypal hot Jupiter, warm Neptune, warm sub-Neptune, and cool super-Earth exoplanets observed using several JWST instruments during both transit and secondary eclipse.  One of their conclusions was that NIRISS transit spectra alone can often constrain the major molecular constituents of clear solar atmospheres, although additional wavelength coverage may be required in certain cases.  \cite{2017ApJ...835...96H} and \cite{2017AJ....153..151B} used information content analysis--often used in studies of solar system atmospheres--to explore optimization of multiple JWST instruments and modes during observations of exoplanet atmospheres in transmission.  In their examination of 11 transiting hot Jupiters, \cite{2017ApJ...835...96H} found that within the constraints of their model, NIRISS consistently provides the most information content for a given integration time.  \cite{2017AJ....153..151B} studied an $R_{\rm pl} = 1.39R_{\rm Jupiter}$ planet of various temperatures, C/O ratios, and metallicities, orbiting WASP-62.  They found that a single observation with NIRISS SOSS always provides the spectra with the highest information content and tightest constraints.  Additionally, when combining two modes, the highest information content spectra with tightest constraints are found by combining NIRISS SOSS with NIRSpec G395 M/H.  

Here, we build upon these previous studies by predicting the properties of the population of TESS discoveries with $R_{\rm pl} \leq 4R_{\oplus}$ that will be most conducive to NIRISS follow-up transit spectroscopy observations, and we then apply our findings to estimate the scope of a JWST follow-up program to map the transition region between hydrogen-dominated and high molecular weight atmospheres.  

\cite{2017arXiv170103539C} also predicted the capabilities of NIRISS, as well as of SOPHIE and SPIRou, in a follow-up program of the \cite{2015ApJ...809...77S} anticipated TESS discoveries.  Our work differs from theirs in that we investigate only NIRISS observations, but we attempt to do so as realistically as possible.  We include the sensitivity of our results to factors such as clouds, planetary composition, observational overhead, and systematic noise.  Our instrument simulator also more closely emulates actual NIRISS observations for all of the planets considered, and we employ theoretical transmission spectra (rather than an atmospheric annulus) to estimate the signal produced by the planetary atmosphere during transit.  In addition, to improve the reliability of our results, we analyze 50 Monte Carlo realizations of the TESS primary mission.  Prior to JWST follow-up observations of TESS-discovered exoplanets, further characterization of TESS planet masses using the radial velocity technique with instruments such as SOPHIE or SPIRou will be required.  In particular, the SPIRou near-infrared spectrometer will be important in characterizing the masses of small planets such as those examined in this work.  Thus, our results are complementary to those of \cite{2017arXiv170103539C} in defining a JWST follow-up strategy for TESS-discovered exoplanet observations.  

Recently, \cite{2017arXiv170800016C} examined characteristics of six well-studied warm Neptunes, or short period planets of size $2R_\oplus < R_{\rm pl} < 6R_\oplus$.  They found that the amplitude of a given planet's spectral transmission features correlates with either the planet equilibrium temperature, or with the bulk mass fraction of H/He in the planetary atmosphere.  \cite{2017arXiv170800016C} applied their findings to the \cite{2015ApJ...809...77S} anticipated TESS discoveries within the same radius regime, estimating the observation time required to distinguish features in planetary spectra using NIRISS.  They show that the number of warm Neptune TESS planets amenable to atmospheric characterization may decrease by up to a factor of eight if transmission amplitude decreases linearly with the bulk mass fraction of H/He.  The work of \cite{2017arXiv170800016C} serves to identify trends in features of TESS planet discoveries that can be used to select the best planets for atmospheric characterization follow-up studies.

Our paper is organized as follows.  In Section \ref{sec:methods}, we describe pertinent exoplanet system properties for both the anticipated TESS discoveries and the existing exoplanets; we discuss our use of stellar and transmission spectra; and we describe the function of our NIRISS simulation tool.  In Section \ref{sec:results}, we present our findings, quantifying the sensitivity of our results to such factors as planetary composition, observational overhead, planetary cloud cover, and existence of systematic noise. We use our estimated S/N to produce simulated spectra for three existing exoplanets.  In Section \ref{sec:JWSTfollowup}, we apply our findings to estimate the scope of a JWST follow-up program devoted to mapping the transition region between high molecular weight and hydrogen-dominated atmospheres.  We summarize in Section \ref{sec:conclusion}.

\section{Methods} \label{sec:methods}

We require three major components to successfully predict NIRISS signal-to-noise.  First, we require system parameters for the planetary systems that we wish to observe.  Below, we describe our use of anticipated TESS discoveries in Section \ref{subsec:TESSdisc}, and our use of planets already discovered in space-based and ground-based surveys in Section \ref{subsec:extantplanets}.  The second major component is model spectra for both the star and the planetary atmosphere, which we describe in Section \ref{subsec:spectra}.  Finally, we require a simulator that models NIRISS operational performance, which we present in Section \ref{subsec:niriss_sim}.   

\subsection{Predicted Properties of TESS Discoveries} \label{subsec:TESSdisc}

\cite{2015ApJ...809...77S} used Monte Carlo simulations to predict the properties of the planets that TESS is likely to detect, and published a catalog of 1,984 planets representing the planet yield from a single Monte Carlo realization of the TESS primary mission. The published catalog only includes planets detected from preselected target star observations rather than full frame image data.  The target star detections make up all of the expected TESS detections for $R_{\rm pl} < 2R_{\oplus}$, and $\sim$30\% of the detections for $2R_{\oplus} < R_{\rm pl} < 4 R_{\oplus}$.  \cite{2015ApJ...809...77S} adopted Kepler planet occurrence rates from \cite{2013ApJ...766...81F} for FGK stars, and from \cite{2015ApJ...807...45D} for stars with $T_{\rm eff} < 4000$ K.  

The \cite{2015ApJ...809...77S} TESS simulated planet catalog contains properties of the planetary systems, which we use as inputs to the NIRISS simulator.  Parameters include stellar temperature and radius, distance and J-band magnitude, and planetary radius and insolation.  We assume all planets are on circular orbits with an impact parameter of $0.5$ during transit.  From the catalog properties, we can calculate other required quantities, such as transit duration ($T_{\rm 14}$) and orbital semi-major axis.  Importantly, the simulated TESS planet catalog does not contain planetary mass, which is required to calculate the scale height of the atmosphere for the planets, and thus is important in estimating the transmission spectroscopy signal.

To calculate the mass of the TESS planets, we explored using a variety of empirical mass-radius relationships \citep{2014ApJ...783L...6W, 2016ApJ...825...19W, 2017ApJ...834...17C}, and ultimately adopted the \cite{2017ApJ...834...17C} model.  \cite{2017ApJ...834...17C} examine 316 objects with well-constrained masses and radii to develop their relationship.  In fitting their empirical data, they employ a broken power-law spanning four regimes which they describe as Terran, Neptunian, Jovian, and Stellar worlds.  Unlike other approaches, in their analysis \cite{2017ApJ...834...17C} treat the transition points between regimes as free parameters.  Their final result is a probabilistic model with credible intervals of values for the transition points, power-law indices in each regime, and radius dispersion in each regime.

In our application of the \cite{2017ApJ...834...17C} broken power-law model to TESS planets with $R_{\rm pl} < 4R_{\oplus}$, we employ only the portion of the model valid for Terran and Neptunian worlds, which is given by

\begin{equation}\label{eq:chenkip1}
M_{\rm pl} = 0.9718R_{\rm pl}^{3.58}\end{equation} 
for $R_{\rm pl} < 1.23 R_\oplus$, and 
\begin{equation}\label{eq:chenkip2}
M_{\rm pl} = 1.436R_{\rm pl}^{1.70}\end{equation} 
for $1.23 R_\oplus \leq R_{\rm pl} < 14.26 R_\oplus$, where mass and radii are given in units of $M_\oplus$ and $R_\oplus$, respectively.  We apply only the basic model and do not vary the parameters within the credible interval.

\begin{figure*}
\gridline{\fig{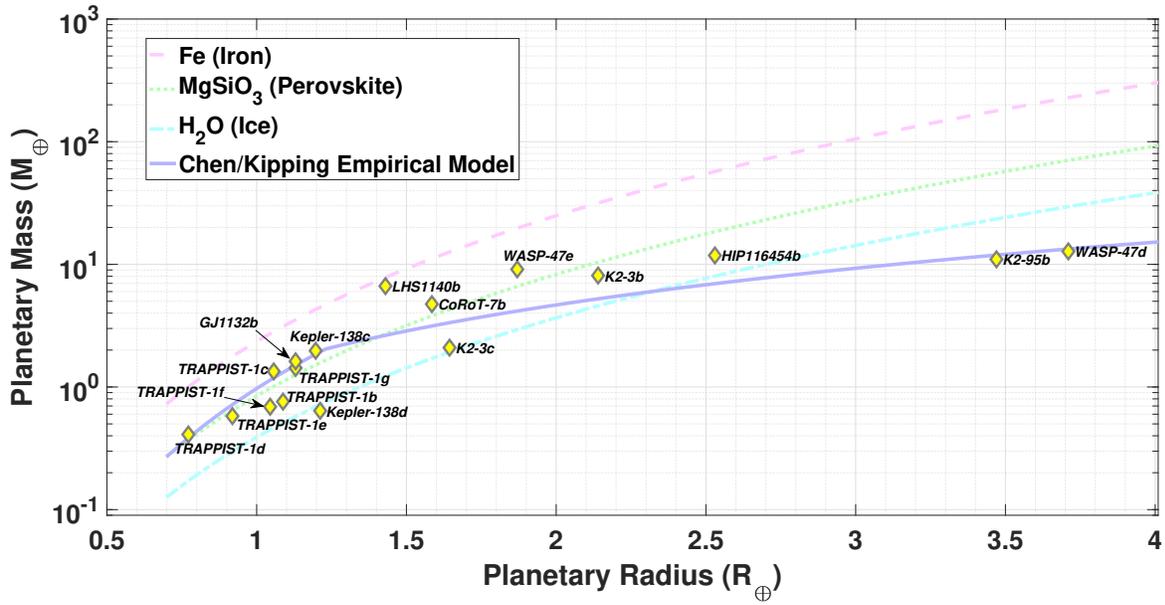}{1.0\textwidth}{(a)}}
\gridline{\fig{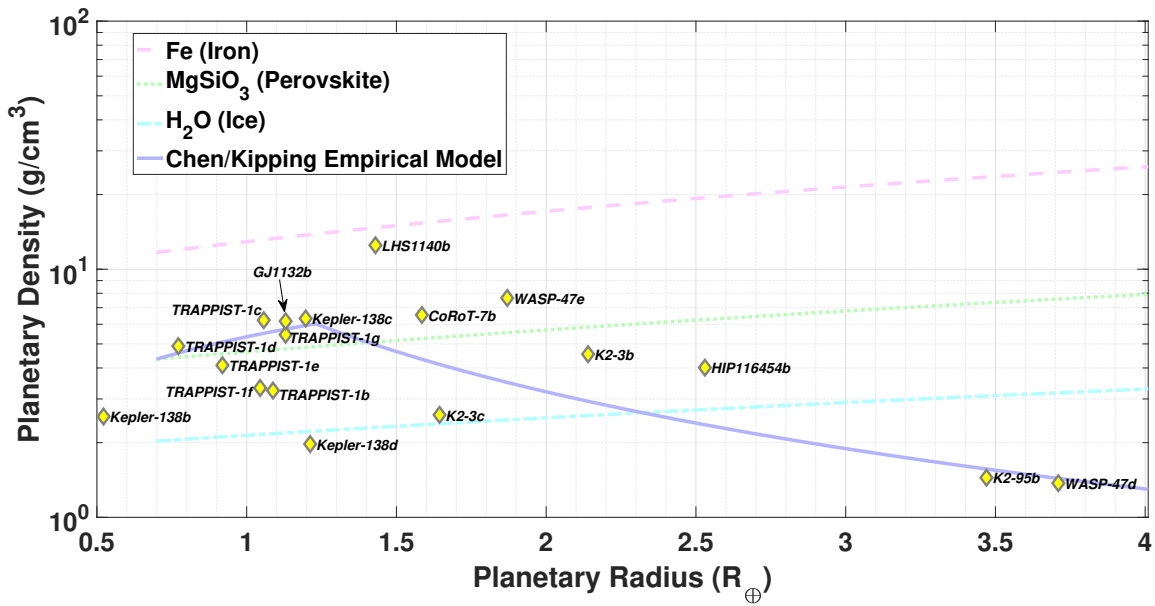}{1.0\textwidth}{(b)}}
\caption{Comparison of the empirical \cite{2017ApJ...834...17C} mass-radius relationship (equations \ref{eq:chenkip1} and \ref{eq:chenkip2}) to the theoretical models of \cite{2007ApJ...669.1279S}.  In (a) we show the masses calculated for each model applied to the \citet{2015ApJ...809...77S} predicted TESS exoplanet discoveries.  In (b) we show the resulting exoplanet densities.  Masses and densities of space-based and ground-based discoveries analyzed in this project are plotted for comparison.\label{fig:comparechenkipmass}}
\end{figure*}

In reality, the planets that TESS discovers will not fall squarely upon the \cite{2017ApJ...834...17C} mass-radius relationship, but will exhibit some variation depending upon planetary composition, which in turn will affect the signal-to-noise (S/N) we compute in this project.  To determine the sensitivity of alternative planetary compositions on our results, we show in Section \ref{sec:results} the S/N attained for planets composed of pure iron (Fe), pervoskite (MgSiO$_3$), and ice (H$_2$O).  In Figure \ref{fig:comparechenkipmass}, we compare the \cite{2017ApJ...834...17C} mass-radius relationship to the theoretical mass-radius relationships found in \cite{2007ApJ...669.1279S} for planets of homogeneous compositions.  All models are applied to the \citet{2015ApJ...809...77S} predicted TESS exoplanet discoveries. 

The \cite{2015ApJ...809...77S} catalog of TESS discoveries represents only one possible realization of the TESS primary mission.  To improve statistical confidence in our results, we also applied our NIRISS simulator to Monte Carlo simulations for 50 trials of the TESS primary mission, provided by \cite{2017arXiv170508891B}. \cite{2017arXiv170508891B} used the same techniques as \cite{2015ApJ...809...77S} in determining the planet yield for each trial of the TESS primary mission.\footnote{Fifty trials of TESS primary mission from \cite{2017arXiv170508891B} may be found at \url{http://scholar.princeton.edu /jwinn/extended-mission-simulations}.} 

\subsection{Selection of Known Exoplanets for Comparison} \label{subsec:extantplanets}

We examined the NASA Exoplanet Archive\footnote{\url{http://exoplanetarchive.ipac.caltech.edu}} \citep{2013PASP..125..989A} and the literature seeking Neptune-sized or smaller ($R_{\rm pl} \leq 4R_\oplus$) confirmed exoplanet discoveries likely to produce strong signals in transmission if observed by NIRISS.  For example, with all other factors the same, an exoplanet with a larger planet-to-star radius ratio will produce a larger transmission spectrum signal.  In addition, a planet with a smaller density will have an atmosphere with a larger scale height, which will also produce a higher signal.  In searching the archive, we sought planets with masses estimated through observations that are orbiting host stars smaller than the Sun.  Tables \ref{tab:spacedisc} and \ref{tab:ground} list the exoplanets and their parameters used in this study.  In Sections \ref{subsubsec:space} and \ref{subsubsec:grounddisc} we discuss further  details concerning our choice of these exoplanets for examination.

\subsubsection{Space-Based Exoplanet Discoveries} \label{subsubsec:space}

We primarily examine space-based discoveries from the K2 mission \citep{2014PASP..126..398H}.  However, we also look at one discovery from the COnvection, ROtation and planetary Transits (CoRoT) minisatellite mission \citep{2006ESASP1306...33B}, as well as one multiplanetary system discovered during Kepler's primary mission \citep{2009IAUS..253..289B}.

In October 2009, \cite{2009A&A...506..287L} announced CoRoT-7b, heralding the discovery as the first super-Earth with a measured radius.  The planet serves as an interesting point of comparison to TESS discoveries, since it orbits a G9V host star that is hotter than most of the best anticipated TESS discoveries.

Kepler discovered thousands of exoplanet candidates, but only some of these planets have estimated masses.  A search of the NASA Exoplanet Archive\footnote{Accessed 3 May 2017} reveals that Kepler-138 is the only M-dwarf hosting planets with estimated masses and with radii less than $1.5R_\oplus$.  We consider this system to determine how the S/N compares to the TESS discoveries.

The K2 mission differs from and is complementary to TESS in that the target fields are located along the ecliptic plane, a field that is not observed during the TESS primary mission.  In this study, we first compare the anticipated TESS-discovered exoplanets to those K2-discoveries with M dwarf host stars.  In particular, we look at two planets from the K2-3 system, as well as K2-95b.  We did not look at K2-3d due to uncertainties in its mass estimate.  We add to this list HIP-116454b, a $2.53R_\oplus$ planet orbiting a bright K-dwarf which was K2's first exoplanet discovery.  The planet serves as an interesting point of comparison because its brightness is near the NIRISS J-band limiting magnitude, and the host star is somewhat hotter than the best-anticipated TESS discoveries.  We also examine Neptune-sized WASP-47d and the super-Earth WASP-47e, which orbit a G9V host star with an estimated radius slightly larger than our Sun's.  Because these latter three planets orbit host stars that are hotter, and thus larger, than many TESS discoveries, we expect the transit depths to be smaller, and thus the S/N to be lower, than most TESS discoveries.

{\catcode`\&=11
\gdef\A{\cite{2009A&A...506..287L}}}

{\catcode`\&=11
\gdef\B{\cite{2014A&A...569A..74B}}}

{\catcode`\&=11
\gdef\C{\cite{2015A&A...581L...7A}}}

\begin{deluxetable*}{cDDDcDDcDD}
\tablecaption{Space-based exoplanet discoveries used for comparison\label{tab:spacedisc}}
\tablecolumns{10}
\tablewidth{0pt}
\tablehead{
\colhead{Exoplanet} & \multicolumn2c{J} & \multicolumn2c{Distance} & \multicolumn2c{Stellar} &
\colhead{Stellar} & \multicolumn2c{Planet} & \multicolumn2c{Planet} &
\colhead{Planet} & \multicolumn2c{Semi-major} & \multicolumn2c{Impact} \\
\colhead{} & \multicolumn2c{} & \multicolumn2c{} & \multicolumn2c{Radius, $R_*$} &
\colhead{Temperature} & \multicolumn2c{Radius, $R_{\rm pl}$} & \multicolumn2c{Mass, $M_{\rm pl}$} &
\colhead{Temperature\tablenotemark{1}} & \multicolumn2c{Axis} & \multicolumn2c{Parameter, $b$} \\
\colhead{} & \multicolumn2c{(mag)} & \multicolumn2c{(pc)} & \multicolumn2c{($R_{\odot}$)} &
\colhead{($ {\rm K}$)} & \multicolumn2c{($R_{\oplus}$)} & \multicolumn2c{($M_{\oplus}$)} &
\colhead{($ {\rm K}$)} & \multicolumn2c{(${\rm AU}$)} & \multicolumn2c{} \\
}
\decimals
\startdata
CoRoT-7b\tablenotemark{2} & 10.301 & 153.7 & 0.820 & 5259 & 1.585 & 4.73 & 1756 & 0.017016 & 0.713 \\
HIP-116454b\tablenotemark{3} & 8.60 & 55.2 & 0.716 & 5089 & 2.53 & 11.82 & 690 & 0.0906 & 0.65 \\
K2-3b\tablenotemark{4} & 9.421 & 42 & 0.561 & 3896 & 2.14 & 8.1 & 463 & 0.0769 & 0.54 \\
K2-3c\tablenotemark{5} & 9.421 & 42 & 0.561 & 3896 & 1.644 & 2.1 & 344 & 0.1405 & 0.31 \\
K2-95b\tablenotemark{6} & 13.312 & 172 & 0.402 & 3471 & 3.47 & 10.99 & 415 & 0.0653 & 0.32 \\
Kepler-138b\tablenotemark{7} & 10.293 & 66.5 & 0.442 & 3841 & 0.522 & 0.066 & 444 & 0.077 & 0.53 \\
Kepler-138c & 10.293 & 66.5 & 0.442 & 3841 & 1.197 & 1.970 & 409 & 0.0906 & 0.922 \\
Kepler-138d & 10.293 & 66.5 & 0.442 & 3841 & 1.212 & 0.640 & 344 & 0.12781 & 0.767 \\
WASP-47d\tablenotemark{8} & 10.613 & 200 & 1.18 & 5475 & 3.71 & 12.75 & 986 & 0.0846 & 0.18 \\
WASP-47e & 10.613 & 200 & 1.18 & 5475 & 1.87 & 9.11 & 2221 & 0.01667 & 0.17 \\
\enddata
\tablenotetext{1}{ Equilibrium temperatures are calculated assuming zero albedo and uniform redistribution of heat.}
\tablenotetext{2}{       CoRoT-7b system parameters from \A, \B, and \cite{2014MNRAS.443.2517H}}
\tablenotetext{3}{       HIP-116454b system parameters from \cite{2015ApJ...800...59V}}
\tablenotetext{4}{       K2-3b system parameters from \cite{2016ApJ...827...78S} and \cite{2016ApJ...823..115D}}
\tablenotetext{5}{       K2-3c system parameters from \cite{2016ApJ...827...78S} and \C}
\tablenotetext{6}{     K2-95b system parameters from \cite{2016AJ....152..223O} and \cite{2017AJ....153...64M}. \cite{2016AJ....152..223O} estimated the planetary mass using an empirical mass-radius relationship.}
\tablenotetext{7}{       Kepler-138a, b, and c system parameters from \cite{2012ApJ...750L..37M}, \cite{2015Natur.522..321J}, and \cite{2017ApJ...835..239S}}
\tablenotetext{8}{       WASP-47d and e system parameters from \cite{2015ApJ...812L..18B} and \cite{2017AJ....153...70S}}
\end{deluxetable*}

\subsubsection{Exoplanets Discovered by Ground-Based Surveys} \label{subsubsec:grounddisc}

The ground-based discoveries examined in this work were found in either the MEarth survey or the TRAnsiting Planets and PlanetesImals Small Telescope (TRAPPIST) survey.  

The MEarth-North and MEarth-South transit surveys were designed to search for super-Earths orbiting mid-to-late M dwarfs with radii less than $0.33 R_\odot$ that are located within 33 parsecs of the Earth \citep{2015csss...18..767I, 2009IAUS..253...37I, 2008PASP..120..317N}.  Here, we examine two MEarth-South discoveries: GJ1132b and LHS1140b.

\cite{2015Natur.527..204B} announced the discovery of GJ1132b in November 2015.  Since that time, further studies conducted using ground-based facilities and the Spitzer Space Telescope have allowed further refinement of system parameters such as stellar and planetary radii \citep{2016arXiv161109848D}.  The water atmosphere we examine in this work is one plausible atmospheric composition for GJ1132b.  In addition, \cite{2017Natur.544..333D} recently reported observations of LHS1140b, a nearby $1.4R_\oplus$ super-Earth orbiting its M dwarf host star within the habitable zone.  The authors note that if this super-Earth had an extended magma-ocean phase, then water may have remained in the mantle until the star reached its current luminosity, thus allowing the presence of water in its atmosphere.

One goal of the TRAPPIST survey is to monitor a select sample of ultra-cool dwarf stars for planetary transits \citep{2011Msngr.145....2J}.  \cite{2016Natur.533..221G, 2017Natur.542..456G} determined that TRAPPIST-1 hosts at least seven Earth-sized planets that are likely to be tidally synchronized.  \cite{2016Natur.537...69D} analyzed the combined transmission spectrum of TRAPPIST-1 b and c, determining that the featureless spectrum ruled out a cloud-free hydrogen-dominated spectrum on the two planets.  However, heavier atmospheres, such as the water atmosphere studied here, remain plausible.  Although here we examine the signal produced by water lines in a water atmosphere, we note that the actual atmospheric composition of the TRAPPIST-1 planets depends upon several factors, such as X-Ray and EUV fluxes, which are areas of active research \citep{2017arXiv170206936O, 2017MNRAS.465L..74W, 2017A&A...599L...3B}.

Multiple recent studies \citep{2017arXiv170402261Q, 2017arXiv170404290W} have estimated the masses of the TRAPPIST-1 planets in order to further constrain the planetary compositions.  In this work, we use the planetary masses inferred by \cite{2017arXiv170402261Q} using N-body dynamical simulations that determine planetary parameters stable over millions of years.

The masses of the TRAPPIST-1 planets and of LHS-1140b are not well constrained.  Since the density of the planet used in simulations impacts the scale height and thus the predicted S/N, uncertain mass estimates may lead to incorrect conclusions.  Thus, for comparison, we also calculate S/N for the TRAPPIST-1 planets and LHS-1140b by assuming an \textit{Earthlike} composition.  We use the semi-empirical mass-radius relationship developed by \cite{2016ApJ...819..127Z}, assuming a core mass fraction of 0.3, the same as that for Earth and Venus.  Since the relationship does not apply to planets with masses less than $1M_\oplus$, for TRAPPIST-1d and TRAPPIST-1e, we simply assume the density is the same as that of Earth's.  Table \ref{tab:Earthlike} shows the masses and densities we used for these alternative calculations.

\begin{deluxetable*}{cDDDcDDcDD}
\tablecaption{Ground-based exoplanet discoveries used for comparison\label{tab:ground}}
\tablecolumns{10}
\tablewidth{0pt}
\tablehead{
\colhead{Exoplanet} & \multicolumn2c{J} & \multicolumn2c{Distance} & \multicolumn2c{Stellar} &
\colhead{Stellar} & \multicolumn2c{Planet} & \multicolumn2c{Planet} &
\colhead{Planet} & \multicolumn2c{Semi-major} & \multicolumn2c{Impact} \\ 
\colhead{} & \multicolumn2c{} & \multicolumn2c{} & \multicolumn2c{Radius, $R_*$} &
\colhead{Temperature} & \multicolumn2c{Radius, $R_{\rm pl}$} & \multicolumn2c{Mass, $M_{\rm pl}$} &
\colhead{Temperature\tablenotemark{1}} & \multicolumn2c{Axis} & \multicolumn2c{Parameter, $b$} \\
\colhead{} & \multicolumn2c{(mag)} & \multicolumn2c{(pc)} & \multicolumn2c{($R_{\odot}$)} &
\colhead{($ {\rm K}$)} & \multicolumn2c{($R_{\oplus}$)} & \multicolumn2c{($M_{\oplus}$)} &
\colhead{($ {\rm K}$)} & \multicolumn2c{(${\rm AU}$)} & \multicolumn2c{} \\
}
\decimals
\startdata
GJ1132\tablenotemark{2} & 9.245 & 12.04 & 0.2105 & 3270 & 1.13 & 1.62 & 579 & 0.01619 & 0.38 \\
LHS-1140b\tablenotemark{3} & 9.612 & 12.47 & 0.186 & 3131 & 1.43 & 6.65 & 230 & 0.0875 & 0.155 \\
TRAPPIST-1b\tablenotemark{4} & 11.4 & 12.1 & 0.117 & 2559 & 1.088 & 0.76 & 400 & 0.01111 & 0.126 \\
TRAPPIST-1c & 11.4 & 12.1 & 0.117 & 2559 & 1.057 & 1.34 & 342 & 0.01522 & 0.161 \\
TRAPPIST-1d & 11.4 & 12.1 & 0.117 & 2559 & 0.722 & 0.41 & 288 & 0.02144 & 0.170 \\
TRAPPIST-1e & 11.4 & 12.1 & 0.117 & 2559 & 0.919 & 0.58 & 251 & 0.02818 & 0.120 \\
TRAPPIST-1f & 11.4 & 12.1 & 0.117 & 2559 & 1.045 & 0.69 & 219 & 0.03707 & 0.382 \\
TRAPPIST-1g & 11.4 & 12.1 & 0.117 & 2559 & 1.130 & 1.43 & 199 & 0.04510 & 0.421 \\
\enddata
\tablenotetext{1}{ Equilibrium temperatures are calculated assuming zero albedo and uniform redistribution of heat.}
\tablenotetext{2}{       GJ1132 system parameters from \cite{2015Natur.527..204B} and \cite{2016arXiv161109848D}}
\tablenotetext{3}{       LHS-1140b system parameters from \cite{2017Natur.544..333D}}
\tablenotetext{4}{       TRAPPIST-1 system parameters from \cite{2017Natur.542..456G} and \cite{2017arXiv170402261Q}}
\end{deluxetable*}

\begin{deluxetable*}{cDD}
\tablecaption{Masses for select exoplanets assuming an Earthlike composition\label{tab:Earthlike}}
\tablecolumns{3}
\tablewidth{0pt}
\tablehead{
\colhead{Exoplanet} & \multicolumn2c{Planet} & \multicolumn2c{Planet} \\ 
\colhead{} & \multicolumn2c{Mass, $M_{\rm pl}$} & \multicolumn2c{Density} \\
\colhead{} & \multicolumn2c{($M_{\oplus}$)} & \multicolumn2c{(g cm$^{-3}$)} \\
}
\decimals
\startdata
LHS-1140b & 3.73 & 7.03 \\
TRAPPIST-1b & 1.36 & 5.81 \\
TRAPPIST-1c & 1.22 & 5.69 \\
TRAPPIST-1d & 0.38 & 5.51 \\
TRAPPIST-1e & 0.78 & 5.51 \\
TRAPPIST-1f & 1.17 & 5.65 \\
TRAPPIST-1g & 1.56 & 5.96 \\
\enddata
\end{deluxetable*}

\subsection{Spectra} \label{subsec:spectra}

The models we use for stellar and transmission spectra allow us to estimate both the signal and the noise that we detect from a given planetary system.  With stellar model spectra, we determine the number of photons output by a host star at each wavelength across the NIRISS bandpass.  The photons received from the host star make up the major source of noise in our simulations.  Our transmission spectroscopy code provides the wavelength-dependent fraction of the stellar area that is blocked by an exoplanet during transit, which constitutes the signal in our simulations.  In Sections \ref{subsubsec:stellar} and \ref{subsubsec:transmission}, we describe these important inputs to our NIRISS simulator.   

\subsubsection{Stellar Spectra} \label{subsubsec:stellar}

PHOENIX/BT-NextGen and PHOENIX/BT-Settl stellar emission spectrum grids \citep{2012RSPTA.370.2765A} provide stellar flux across the NIRISS bandpass at a resolution of $\sim$2 Angstroms.  We employ solar metallicity spectra, which include the effects of absorption from molecules such as water vapor \citep{2006MNRAS.368.1087B} in the stellar atmosphere.  The host stars of our target systems range in effective temperature from 2,090 $ {\rm K}$ to 14,655 $ {\rm K}$.  We employ BT-Settl models for host stars with effective temperatures less than 2,600 ${\rm K},$\footnote{In all of our calculations, only 43 systems adopt BT-Settl stellar models.  For the published \cite{2015ApJ...809...77S} planetary systems, only one has a host star of temperature less than 2,600 ${\rm K}$, and for the 50 Monte Carlo realizations of the TESS primary mission, only 42 stellar systems (out of 124,173) have host stars with effective temperatures less than 2,600 ${\rm K}$.} and BT-NextGen models for those systems with higher effective temperatures.  BT-NextGen models span stellar temperature values from 2,600 $ {\rm K} $ to 70,000 $ {\rm K}$, and $\log(g)$ values from $-0.5$ to $6.0$.  Stellar models are provided in $\log(g)$ increments of 0.5, and in temperature increments of 100 $ {\rm K}$ up to a stellar temperature of 7,000 $ {\rm K}$, then in temperature increments of 200 ${\rm K}$ up to a temperature of 12,000 $ {\rm K}$, and thereafter in temperature increments of 500 $ {\rm K}$.  Our simulation routine selects the stellar model with effective temperature and $\log(g)$ values closest to the particular planetary system we wish to observe.  

\subsubsection{Planetary Atmospheres and Transmission Spectra} \label{subsubsec:transmission}

The TESS planet catalog discussed in Section \ref{subsec:TESSdisc} includes planetary radii, but the radii do not vary with wavelength, and the \cite{2015ApJ...809...77S} simulations make no assumptions regarding planetary atmospheres.  For the known planets discussed in Section \ref{subsec:extantplanets}, we can make some assumptions regarding likely planetary atmospheres based upon observations and estimated bulk densities.  Recent research has identified the radius regime from $1.5R_\oplus$ to $2R_\oplus$ as the transition region from rocky, terrestrial planets with high molecular weight atmospheres to sub-Neptune planets enveloped in hydrogen-dominated atmospheres \citep{2008ApJ...685.1237E,2009ApJ...690.1056M,2011IAUS..276..212K,2014ApJS..210...20M,2015ApJ...801...41R,2017arXiv170310375F}.

\textit{Clear Atmospheres--}For the purposes of our calculations, we assume that \textit{all} planets with $R_{\rm pl} \leq 1.5R_\oplus$ are enveloped in a clear, isothermal water atmosphere (mean molecular weight $\mu = 18$), and we assume that \textit{all} planets with $1.5R_\oplus < R_{\rm pl} \leq 4R_\oplus$ are surrounded by a clear, isothermal hydrogen-dominated atmosphere ($\mu = 2.39$). We use an isothermal atmosphere because we do not have sufficient information to justify a more complex structure, and also because transit spectra are not directly sensitive to the source function in the exoplanetary atmosphere. For each planet, we calculate the equilibrium temperature assuming zero albedo and uniform redistribution of heat.  These temperatures are reported for known exoplanets in Tables \ref{tab:spacedisc} and \ref{tab:ground}.  In reality, some planets in the transition region will likely have heavy atmospheres, some will likely have lighter atmospheres, and some will be rocky cores stripped of an atmosphere.  By assuming a single atmospheric composition within the two radii regimes, we can better determine the contributions of other planetary parameters on resultant S/N.

For most planetary compositions, the spectrum of water vapor will dominate the NIRISS spectral region.  We estimate the signal produced by water lines in the isothermal planetary atmospheres by employing a version of a spectral transmittance code presented previously in \cite{2013ApJ...774...95D}.  Here, we modify the code to cover the NIRISS bandpass.  Our code uses a slant-path geometry, assuming a 200-layer atmosphere in hydrostatic equilibrium, with pressures in the layers equally spaced in log from 1 to $10^7$ dynes cm$^{-2}$.  The hydrogen-dominated atmospheres include continuous opacity due to collision-induced absorption of H$_2$ \citep{1995ApJ...441..960Z,2002A&A...390..779B}.  For the water atmospheres, we place a solid surface at 1 bar ($10^6$ dynes cm$^{-2}$).  We calculate water opacity using water lines \citep{2006MNRAS.368.1087B} downloaded from the Exomol Database\footnote{\url{http://www.exomol.com/data/data-types/linelist/H2O/1H2-16O/BT2/}, accessed 29 Jan 2017} \citep{2012MNRAS.425...21T}.  Our code scales the strengths of the lines at the isothermal temperature of the planet, and then bins the scaled strengths of the water lines into bins of width 0.01 cm$^{-1}$, much smaller than the NIRISS resolution \citep{2017ApJ...841L...3D}.  We convolve the high-resolution transmission spectrum output by our code to the resolution of the NIRISS instrument prior to employing the output spectrum in our NIRISS simulator.

\begin{figure*}
\gridline{\fig{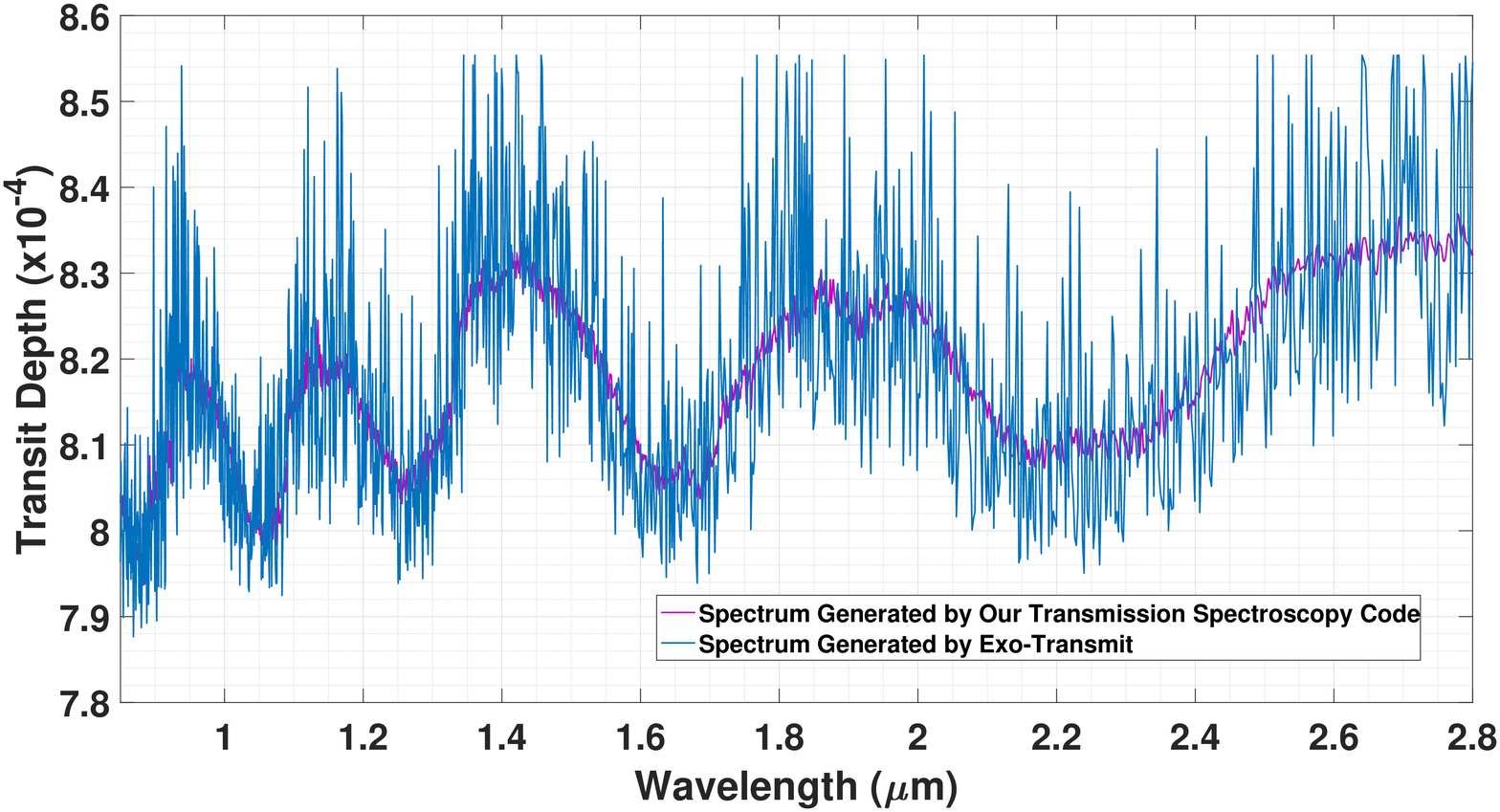}{1.0\textwidth}{}}
\caption{Comparison of our transmission spectroscopy code to Exo-Transmit \citep{2017PASP..129d4402K} for a super-Earth ($R_{\rm pl} = 1.28R_\oplus$) at equilibrium temperature 788 $ {\rm K}$ with a clear water atmosphere.  The system parameters of the chosen super-Earth correspond to mean values of system parameters for the published \cite{2015ApJ...809...77S} TESS planets in the radius regime $R_{\rm pl} \leq 1.5R_{\oplus}$.  Both spectra have resolving powers of 1000 but differ in details due to binning (see text).  \label{fig:transcompare}}
\end{figure*}

We previously validated the code as described in \cite{2013ApJ...774...95D}.  In addition, our code is in close agreement with output results from \cite{2013ApJ...778..183L}, as shown in their Figure 5.  In this work, we further compare our code to the output results of Exo-Transmit \citep{2017PASP..129d4402K}, as shown in Figure \ref{fig:transcompare} for a
super-Earth ($R_{\rm pl} = 1.28R_\oplus$) at equilibrium temperature 788 $ {\rm K}$ with a clear water atmosphere.  The system parameters of the chosen super-Earth correspond to mean values of system parameters for the published \cite{2015ApJ...809...77S} TESS planets in the radius regime $R_{\rm pl} \leq 1.5R_{\oplus}$.  Exo-Transmit produces transmission spectra at a fixed spectral resolution R = 1000.  Our code produces a high-resolution spectrum with millions of lines across the NIRISS bandpass at sub-Doppler resolution (R $>$ 300,000 across the NIRISS bandpass), which we then convolve with a Gaussian to produce the R = 1000 spectrum illustrated in Figure \ref{fig:transcompare}.  The Exo-Transmit spectrum retains more structure since the spectrum is produced at a native resolution R = 1000, whereas our high-resolution code is smoothed to the same resolution.  The average values across the two spectra are in close agreement.  

\textit{Cloud Effects--}Our spectral transmittance code allows placement of an opaque cloud layer at a pressure level of our choice.  In this work, we examine the effects of clouds by placing an opaque cloud deck at a pressure of 10 mbar for planets with both water and hydrogen-dominated atmospheres.

\textit{Use of Transmittance Code with Planetary Systems--}When applying our spectral transmittance code to the known exoplanets, we apply the code to each planet individually, using the system parameters reported in Tables \ref{tab:spacedisc}, \ref{tab:ground}, and \ref{tab:Earthlike}.  However, when applying the code to the \cite{2015ApJ...809...77S} anticipated TESS planets and to the 50 Monte Carlo realizations of the TESS primary mission data, it is computationally expedient to scale the transmission spectra to planetary systems with mean values of stellar radius and planetary radius, mass, and temperature for the \cite{2015ApJ...809...77S} anticipated TESS planets within each radius regime.  The remainder of this subsection outlines the method we employ in scaling the transmission spectrum of each TESS planet to one of the two reference planets.

Our code outputs a transmission spectrum equivalent to the transit depth of each reference planet.  In general,

\begin{equation} \label{eq:trandep}
{\rm Transit\ Depth} = \frac{\rm Area_{\rm pl}}{\rm Area_{*}} = \frac{(R_{\rm pl}+h)^2}{R_{*}^2} , 
\end{equation}
where $R_{\rm pl}$ is the radius of the solid surface of the planet, $h$ is the wavelength-dependent thickness of the atmosphere, and $R_*$ is the stellar radius.  Expanding equation \ref{eq:trandep} and ignoring small terms, we find
\begin{equation} \label{eq:exptd}
{\rm Transit\ Depth} \approx \frac{R_{\rm pl}^2}{R_{*}^2} + \frac{2R_{\rm pl}h}{R_{*}^2}.
\end{equation}
The second term in equation \ref{eq:exptd} is the wavelength-dependent term that must be scaled to each planetary system.  We note that since $h$ represents the thickness of the atmosphere, it must be proportional to the atmospheric scale height $H$, which is found from 
\begin{equation} \label{eq:H}
H = \frac{kT_{\rm pl}}{\mu g} = \frac{kT_{\rm pl}R_{\rm pl}^2}{\mu GM_{\rm pl}},
\end{equation}
where $k$ is Boltzmann's constant, $T_{\rm pl}$ is the equilibrium temperature of the planet, and $g=\frac{GM_{\rm pl}}{R_{\rm pl}^2}$ is the acceleration due to gravity on the planet, where $G$ is the universal gravitational constant and $M_{\rm pl}$ is the mass of the planet.  Denoting the second term in equation \ref{eq:exptd} with the variable S, substituting equation \ref{eq:H} for $H \propto h$, and dropping various constants, we find
\begin{equation} \label{eq:S}
S = \frac{2R_{\rm pl}h}{R_{*}^2} \propto \frac{T_{\rm pl}R_{\rm pl}^3}{M_{\rm pl}R_{*}^2} \propto \frac{T_{\rm pl}}{\rho_{\rm pl}R_{*}^2} ,
\end{equation}
where $\rho_{\rm pl}$ is the density of the planet.

We use equation \ref{eq:S} to scale the atmosphere of any TESS planet in one of the two radius regimes to the appropriate reference planet.  Specifically, using the subscript $T$ to refer to the TESS planetary system, and the subscript ``{\rm ref}'' to refer to the reference planetary system, we have
\begin{equation} \label{eq:scaling}
S_{\rm T} = S_{\rm ref}\times \frac{T_{\rm pl,T}}{T_{\rm pl,ref}}\times \frac{\rho_{\rm pl,ref}}{\rho_{\rm pl,T}}\times \frac{R_{\rm *,ref}^2}{R_{\rm *,T}^2},
\end{equation}
where $S_{\rm T}$ is the wavelength-dependent transmission spectrum for the TESS system, scaled to the reference transmission spectrum $S_{\rm ref}$ for the regime of interest.

\subsection{Description of NIRISS Simulator} \label{subsec:niriss_sim}

In this work, our simulator predicts the S/N attainable by NIRISS, operating in SOSS mode, during 10-hour observation programs of the \cite{2015ApJ...809...77S} anticipated TESS discoveries, predicted TESS discoveries from 50 Monte Carlo realizations of the TESS primary mission, and the known exoplanets listed in Tables \ref{tab:spacedisc} and \ref{tab:ground}.  We chose 10 hours per target to represent the order of magnitude observation time, $t_{\rm obs}$,  required to complete a statistical survey project of multiple exoplanets.  Except as noted during our description of observational overhead, our simulations equally apportion the observation time to periods in and out of transit. We do not take into account minor effects such as stellar limb darkening, and we assume that the entire cross sectional area of the planet blocks the star throughout the transit.  For comparison, we also examine the S/N attainable in more intensive 100-hour observation programs of the potentially habitable TESS planets--those with $R_{\rm pl} \leq 1.5R_{\oplus}$ and cool or temperate equilibrium temperature.\footnote{In planning actual observation programs, visibility of the targeted planetary systems and potential contamination by partly overlapping spectra from nearby stars must be considered: \url{http://maestria.astro.umontreal.ca/niriss/SOSS_cont/SOSScontam.php}.  These topics are beyond the scope of this work.} 

We have developed our NIRISS simulator in consultation with the NIRISS instrument team, and we have incorporated the latest NIRISS design values into our simulator to the extent possible.\footnote{\url{http://jwst.astro.umontreal.ca/?page_id=51}}  In this section, we describe in detail the method we use to model NIRISS observations.  We close the section by briefly explaining how we explore the effects of observational overhead and the presence of systematic noise on our S/N results.  

\subsubsection{Predicting JWST/NIRISS S/N} \label{subsubsec:NIRISSSNR}

The NIRISS SOSS mode offers three readout modes covering wavelengths from 0.6 to 2.8 microns \citep{2012SPIE.8442E..2RD}.  Here, we estimate NIRISS S/N only for Order 1 spectra produced in \textit{nominal} (256 x 2048 pixel subarray) and \textit{bright} (96 x 2048 pixel subarray) modes, spanning wavelengths from 0.8 to 2.8 microns.\footnote{The blue cut-off in \textit{bright} mode is 0.9 microns due to the smaller size of the subarray.}  The \textit{bright} mode read time (2.213 sec) is faster than that of \textit{nominal} mode (5.491 sec), thus allowing observation of brighter targets before saturating the subarray pixels. A weak cylindrical lens at the entrance to the NIRISS cross-dispersed grism spreads light across approximately 23 pixels in the spatial direction, also enabling observation of brighter objects before reaching pixel saturation.  

Given a targeted planetary system for NIRISS observations, our simulation begins by selecting the appropriate PHOENIX stellar model, which provides output stellar flux in ergs sec$^{-1}$ cm$^{-2}$ \AA$^{-1}$.  We convolve the stellar spectrum with a Gaussian corresponding to the NIRISS optics resolution ($\sim$1.6 pixels FWHM), and then convert the model flux output at the star to a photon flux arriving at the JWST observatory by scaling the stellar model to the J-Band magnitude of the star, multiplying by the area of JWST (25.3 m$^2$), and then dividing by the energy per photon $h\nu$, where $h$ is Planck's constant and $\nu$ is the frequency of the photon.  We calculate the signal produced by the planetary atmosphere by determining the wavelength-dependent portion of the stellar photons that are blocked by the atmospheric annulus during transit (equation \ref{eq:trandep}). 

To determine the signal and noise detected by NIRISS, we must account for instrument throughput and dispersion.  The NIRISS design team has provided an estimate of throughput across the NIRISS bandpass at a resolution of 1 nm.  The estimate includes detector quantum efficiency and transmission through all optical elements of the JWST observatory, NIRISS instrument, and GR700XD grism.  We apply the wavelength-dependent throughput to the stellar model and to the wavelength-dependent signal produced by the atmosphere during transit.  The dispersion of 0.974 nm/pixel allows us to determine the wavelength range of each NIRISS subarray column in the dispersion direction.  Knowing the wavelength (and frequency) range covered by each column in the dispersion direction, we can then determine the photon flux spread across the spatial direction both in and out of transit.  Our simulation also incorporates noise due to zodiacal light \citep{1998ApJ...508...44K} and JWST telescope thermal background \citep{2004SPIE.5487..785S}, but these effects are negligible in practice for the observation of the bright transiting planet host stars.  Read noise and dark currents are also regarded as negligible for well-exposed integrations.

Before calculating the signal and noise produced during an observation program, we must first determine the efficiency of the observation.  The on-sky efficiency achievable with NIRISS depends upon the brightness of the targeted stellar system.  The integration time of a given observation is determined by the number of \textit{reads}, also known as $n_{\rm groups}$, performed before resetting the well \citep{2014PASP..126.1134B,2017arXiv170201820B}.  The time to reset the well is equal to the time to perform one read, which we denote as $t_{\rm frame}$.  We calculate integration time\footnote{\url{http://maestria.astro.umontreal.ca/niriss/simu1D/SOSS_Simulator_Guide.pdf}} with
\begin{equation} \label{eq:tint}
t_{\rm int} = t_{\rm frame} \times (n_{\rm groups}-1),
\end{equation}
and on-sky efficiency using
\begin{equation} \label{eq:effic}
\eta = \frac{n_{\rm groups}-1}{n_{\rm groups}+1}.
\end{equation}
Our simulation calculates and employs the maximum on-sky efficiency possible without saturating any pixels.  We do this by calculating the maximum number of $n_{\rm groups}$ we can use without any pixels accumulating greater than the full well capacity of 72,000 electrons.\footnote{We used full well capacity in our calculations since that is the value used by the NIRISS 1D SOSS simulator.  Going to full well capacity could result in systematic effects \citep{2014ApJ...783..113W}.  Thus, we examined using 55,000 electrons rather than full well capacity and found that on-sky efficiency and therefore S/N changes minimally in most cases.  The greatest change in S/N of $\sim$20\% occurs for systems that are viewed at $n_{\rm groups} = 3$ ($\eta = 0.5$) for 72,000 electrons and then $n_{\rm groups} = 2$ ($\eta = 0.33$) for 55,000 electrons.  The conclusions of this work are unaltered by changing the electron level used for saturation.}  In performing this calculation, we make use of the fact that when the photon flux is spread across the pixels in the spatial direction, 7\% of the photons fall in the peak pixel.  In calculating efficiency, we consider only correlated double sampling, where flux is calculated by subtracting the last read from the first read.  The NIRISS design also allows a superbias subtraction method to calculate flux when observing brighter objects, but this mode has more uncertainty in its noise properties, and we do not consider it in this work.

In calculating $n_{\rm groups}$, our simulator first assumes that we are observing in \textit{nominal} mode.  If the required value of $n_{\rm groups}$ is less than 2, which is the minimum value of $n_{\rm groups}$ required when using correlated double sampling, then we recalculate $n_{\rm groups}$ using \textit{bright} mode instead.  For some stellar systems, pixels will saturate even in \textit{bright} mode.  For the \cite{2015ApJ...809...77S} planets, the plot of our output results shows those systems which were observed using \textit{bright} mode, and we also indicate those systems where some pixels saturate.  When analyzing the 50 trials of TESS data, we present our results as a 2-dimensional histogram in S/N-$R_\oplus$ space.  In presenting these results, we ignore those systems where pixel saturation occurs in \textit{bright} mode.\footnote{If the systems where pixel saturation occurs are included, our results support the same conclusions.}  For those systems that are dim, the highest value of $n_{\rm groups}$ our simulator employs is 88 \citep{2014PASP..126.1134B}.  After the optimum number of $n_{\rm groups}$ has been calculated, our simulator determines the on-sky efficiency using equation \ref{eq:effic}.  

Knowing the efficiency, the observation time, and the photon flux from the star falling across the pixels, the stellar shot noise is determined by using
\begin{equation} \label{eq:stnoise}
N_{\rm shot} = \sqrt[]{\eta \ t_{\rm obs} \ F_{*}},
\end{equation}
where $N_{\rm shot}$ represents the photon shot noise and $F_{*}$ is the photon flux falling across the pixels.  Photon shot noise is the primary source of noise for the S/N values we report in Section \ref{sec:results}.  

A similar calculation is possible for the signal.  Knowing the number of stellar photons blocked by the atmospheric annulus across the NIRISS bandpass, we can calculate the signal produced by the water lines in the atmosphere from
\begin{equation} \label{eq:atmsig}
S_{\rm atm} = \eta \ \frac{t_{\rm obs}}{2} \ F_{\rm atm},
\end{equation}
where $S_{\rm atm}$ represents the signal produced by the atmosphere and $F_{\rm atm}$ is that portion of the stellar flux blocked by the atmospheric annulus.

\begin{figure*}
\gridline{\fig{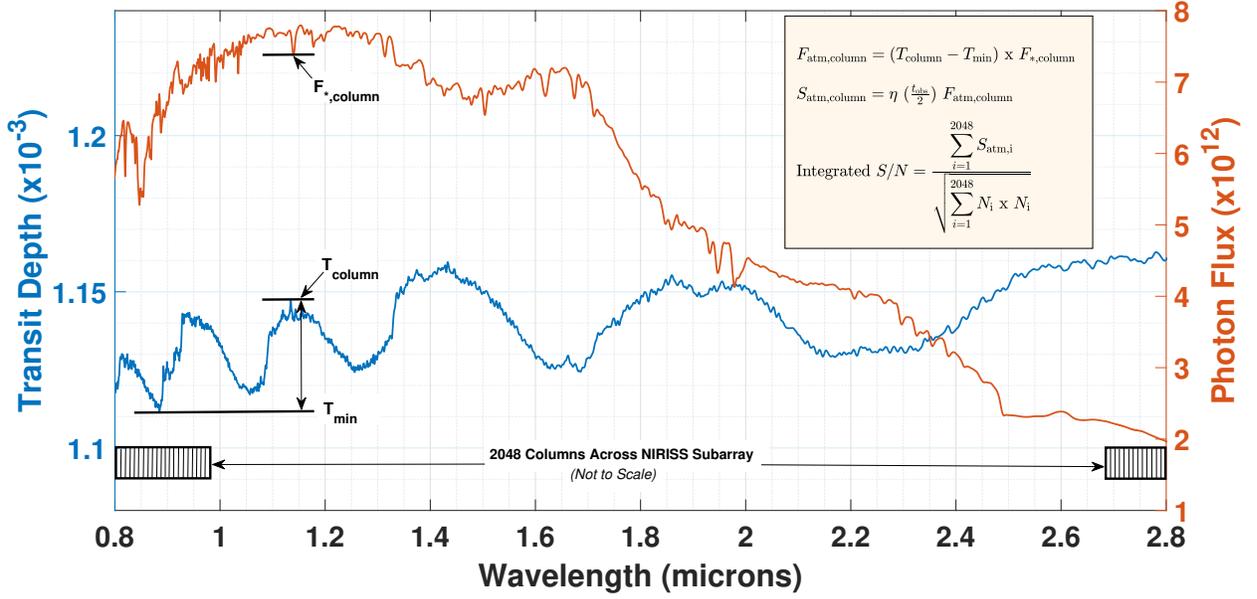}{1.0\textwidth}{}}
\caption{Graphical illustration showing our calculation of signal (equation \ref{eq:atmsig}) in each column of the NIRISS subarray, as well as the integrated S/N across the NIRISS bandpass.  The left vertical axis (blue curve) shows the transit depth, while the right vertical axis (orange curve) shows photon flux.  In calculating integrated S/N, $N_{\rm i}$  is the noise both in and out of transit due to stellar photons, zodiacal light, and JWST telescope background. \label{fig:GraphicalSignal}}
\end{figure*}

The S/N varies greatly across the NIRISS bandpass, which is to be expected since the strength of water lines and the brightness of the host star differ with wavelength.  In this work, after calculating the wavelength-dependent values of $N_{\rm shot}$ and $S_{\rm atm}$ across the columns of the NIRISS subarray, we next calculate the \textit{integrated} S/N for the detection of the atmosphere across the NIRISS bandpass.  In Figure \ref{fig:GraphicalSignal}, we illustrate graphically our calculation of the signal (equation \ref{eq:atmsig}) in each column across the subarray, as well as the integrated S/N, which is what we report on our Figures in Section \ref{sec:results}.

\begin{figure*}
\gridline{\fig{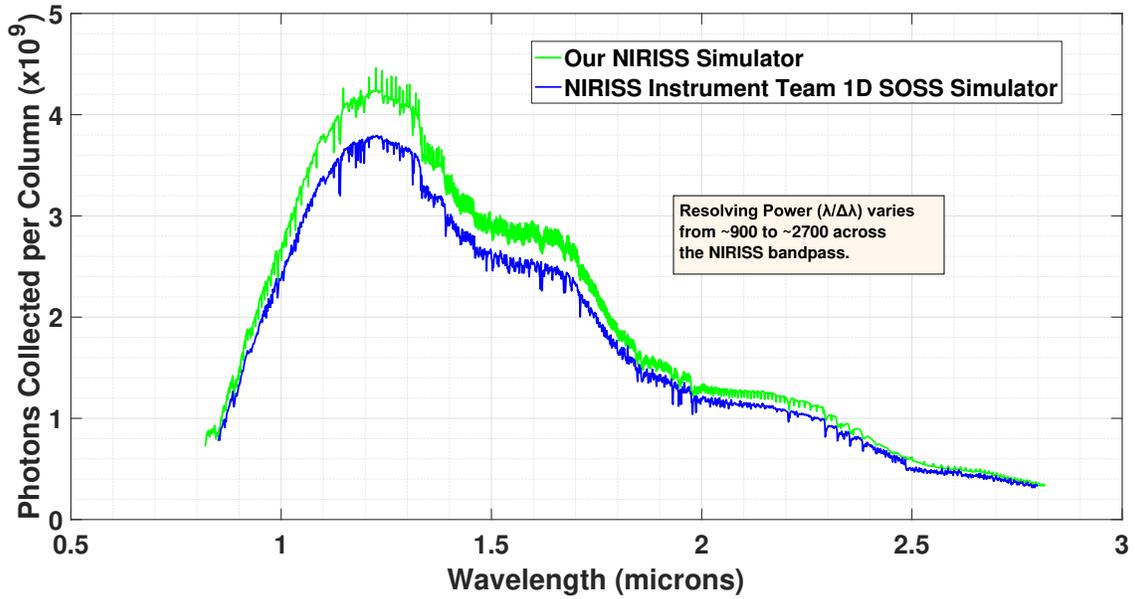}{1.0\textwidth}{}}
\caption{Comparison of the out-of-transit photon count per column of our NIRISS simulator to that of the NIRISS 1D SOSS simulator produced by the instrument design team.  Here, we compare photon counts for 72 transit observations of GJ1132b.  In this case, our photon count per column exceeds that found by the NIRISS 1D SOSS simulator by $\sim$11\%.  For all planetary systems where we compared the two codes, our results differed from those of the 1D SOSS simulator by 8 to 11\%. \label{fig:NIRISSSimcompare}}
\end{figure*}

We validated our NIRISS simulation code by comparing it to the 1D SOSS simulator developed by the NIRISS instrument design team.\footnote{\url{http://maestria.astro.umontreal.ca/niriss/simu1D/simu1D.php}}  For all planetary systems where we compared the two codes, our results differed from those of the 1D SOSS simulator by 8 to 11\%.  In Figure \ref{fig:NIRISSSimcompare}, we depict the out-of-transit stellar photons collected per column on the NIRISS subarray as estimated by both our simulator and that of the NIRISS instrument team for a 100-hour observation program of GJ1132.  In this case, our photon count per pixel exceeds that found by the NIRISS 1D SOSS simulator by $\sim$11\%.

\subsubsection{Observational Overhead} \label{subsubsec:obsovrhd}

Observational overhead accounts for the fact that during actual JWST observations, not all of the telescope time devoted to a program will be used for science.  Rather, some of the time will be used to slew the observatory or set up the instrument.  In general, the clock time for a JWST observation is given by\footnote{\url{https://jwst-docs.stsci.edu/display/JPPOM/Overhead+Duration+Components}}
\begin{equation} \label{eq:tclock}
t_{\rm clock} = (t_{\rm slew} + t_{\rm science} + t_{\rm instrument})\times 1.16 + t_{\rm scheduling}.
\end{equation}
Here, $t_{\rm slew} = 30$ min is the time to slew the telescope;\footnote{The time to slew JWST varies depending upon the slew distance in arcseconds: \url{https://jwst-docs.stsci.edu/display/JPPOM/Slew+Times}.  The longest slew time reported, for 180 degrees, is $\sim$1 hour, so we use half of this time for all TESS systems.} $t_{\rm science}$ is the observation time, both in and out of transit; $t_{\rm instrument} = 17.1$ min is an instrument overhead time;\footnote{Instrument overhead time includes target acquisition (10 minutes) and filter wheel movements.  Further details can be found at: \url{https://jwst-docs.stsci.edu/display/JPPOM/NIRISS+Overheads.}}  and $t_{\rm scheduling} = 60$ min is the additional JWST time required when scheduling an observation with a start time more precise than 24 hours, certainly the case with exoplanet transits.  The factor of 1.16 is observatory overhead that takes into account losses due to calibrations and dead time of JWST.  We note that equation \ref{eq:tclock} is the current expression used to estimate clock time, but the expression is likely to change when JWST is operational.

To determine the effects of observational overhead, we modify equations \ref{eq:stnoise} and \ref{eq:atmsig} so that we calculate $N_{\rm shot}$ and $S_{\rm atm}$ using the number of transits, $n_{\rm trans}$, and transit duration ($T_{\rm 14}$), $t_{\rm trans}$, as
\begin{equation} \label{eq:stnoiseOH}
N_{\rm shot} = \sqrt[]{2 \ \eta \ n_{\rm trans} \ t_{\rm trans} \ F_{*}},
\end{equation}
and
\begin{equation} \label{eq:atmsigOH}
S_{\rm atm} = \eta \ n_{\rm trans} \ t_{\rm trans} \ F_{\rm atm}.
\end{equation}
Then, we calculate $n_{\rm trans}$ using
\begin{equation} \label{eq:ntranOH}
n_{\rm trans} = \frac{t_{\rm obs}}{(2 t_{\rm trans} +  47.1) \times 1.16 + 60},
\end{equation}
where times are in minutes.  As before, we use $t_{\rm obs} = 10$ hrs $= 600$ min.  Note that equations \ref{eq:stnoiseOH} and \ref{eq:atmsigOH} could be used in place of equations \ref{eq:stnoise} and \ref{eq:atmsig} in the simple observation program described previously, where we apportion equal amounts of time in and out of transit, but in that case, the number of transits would be calculated using the simple relationship $n_{\rm trans} = t_{\rm obs}/(2t_{\rm trans})$.

\subsubsection{Systematic Noise per Transit Observation} \label{subsubsec:noisefloor}

In Section \ref{sec:results}, we examine the effects of the systematic noise per transit observation on TESS discoveries, and then go on to show the spectra that may be anticipated for NIRISS observation programs of the known exoplanet K2-3c, both with and without considering this systematic noise.  A relatively low residual systematic noise level is anticipated for JWST observations, since with existing instruments we have already achieved residual noise levels of $\sim$25 ppm with Hubble \citep{2014Natur.505...69K} and $\sim$30 ppm using Spitzer \citep{2016MNRAS.455.2018D}.  Here we adopt the noise floor used by \cite{2016ApJ...817...17G} of 20 ppm for one planetary transit.  We note that the exact nature of the systematic noise will not be known until JWST commissioning. However, we expect that any residual noise will be due to systematic effects that can be represented by an equivalent sine wave because instrumental noise is commonly bandwidth-filtered.  For the observation programs we examine, multiple transits will be observed, with the state of the instrument different for each telescope pointing, and where the equivalent sine wave representing systematic effects is most likely observed at a different phase during each observation.  Thus, we assume the systematic noise will decrease as the square root of the number of independent measurements, or transits.  

To determine the effects of systematic noise on the integrated S/N across the entire NIRISS bandpass for the \cite{2015ApJ...809...77S} anticipated TESS planets, we apply the systematic noise to each resolution element (i.e., 2 columns).  However, in presenting spectra for K2-3c, we bin the S/N into 64 bins across the NIRISS bandpass.  Binning increases S/N above that of the individual NIRISS resolution elements, but reduces resolving power across the bandpass.  For 64 bins, resolving power ranges from almost 30 at the blue cut-off to about 95 at the red cut-off.  For the spectrum where systematic noise has been incorporated, the noise in each of the 64 bins is calculated with

\begin{equation} \label{eq:noisefloor}
  N_{\rm bin, total} = \sqrt[]{N_{\rm bin, shot}^2 \ + \ (\frac{20 \times 10^{-6}}{\sqrt[]{n_{\rm trans}}} \times F_{\rm bin,*})^2},
\end{equation}
where the second term inside the square root on the right-hand side of the equation represents the adjustment made due to systematic noise.

\section{Results and Discussion} \label{sec:results}

In this section, we present our analysis of attainable NIRISS S/N in 10-hour observation programs of the \cite{2015ApJ...809...77S} predicted TESS planets, as well as our analysis of 50 Monte Carlo realizations of TESS primary mission data, with comparisons to expected S/N for known exoplanets.  For comparison, we also examine the S/N attainable in more intensive 100-hour observation programs of the potentially habitable \cite{2015ApJ...809...77S} predicted TESS planets--those with $R_{\rm pl} \leq 1.5R_{\oplus}$ and cool or temperate equilibrium temperatures.  First we show our results for the \cite{2015ApJ...809...77S} planets where mass is calculated using the \cite{2017ApJ...834...17C} mass model.  We then explore the effects of planetary composition, observational overhead, and existence of planetary clouds on these results.  We next show our results from analyzing 50 Monte Carlo realizations of TESS primary mission data.  We conclude the section by presenting our analysis of the effects of systematic noise on our predicted S/N for the TESS anticipated discoveries.  Additionally, based upon our estimates of NIRISS S/N, we create sample spectra for the known exoplanet K2-3c, illustrating the effects of adding systematic noise to a simulated spectrum.  

\subsection{S/N for Anticipated TESS Discoveries} \label{subsec:resultsSNRsullivan}

\begin{figure*}[ht!]
\gridline{\fig{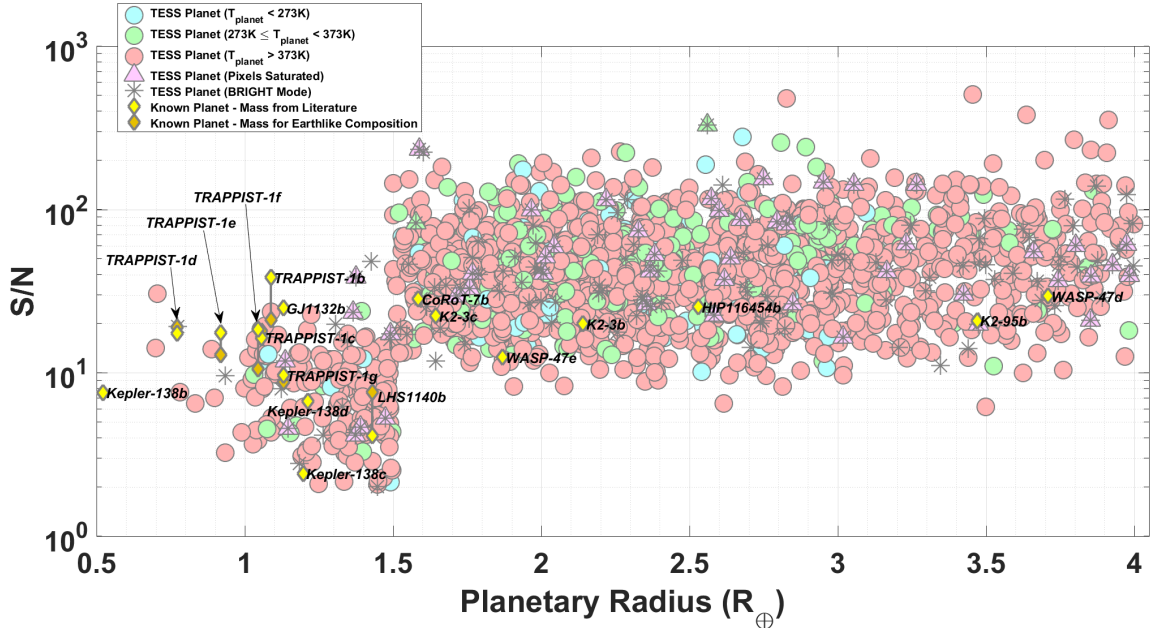}{1.0\textwidth}{}}
\caption{Integrated Signal-to-Noise (S/N) we predict for NIRISS detection of the atmosphere of anticipated TESS planets in 10-hour observation programs of all planets.  We also show integrated S/N for known exoplanets discovered from space-based and ground-based surveys.  Our results show that TESS is likely to find many planets with promising properties in the radius regime $1.5R_\oplus < R_{\rm pl} < 4R_\oplus$.  In particular, the planets found in the radius regime $1.5R_\oplus < R_{\rm pl} < 2R_\oplus$ will help to define the parameter space where planetary atmospheres transition from hydrogen-dominated to high molecular weights.  However, TESS is unlikely to discover many terrestrial-sized planets more amenable to atmospheric characterization than those that have already been discovered.  We emphasize that the outcome for terrestrial-sized planets is based upon Kepler-derived occurrence rates, and that co-planar compact multi-planet systems (e.g., TRAPPIST-1) may be under-represented in the predicted TESS planet yield. The apparent step function at 1.5$R_{\oplus}$ results from assuming all planets with $R_{\rm pl} \leq 1.5R_{\oplus}$ have water atmospheres, and all planets with $1.5R_{\oplus} < R_{\rm pl} < 4R_{\oplus}$ have hydrogen-dominated atmospheres (Section \ref{subsubsec:transmission}).  \label{fig:sulldata}}
\end{figure*}

\begin{figure*}[ht!]
\gridline{\fig{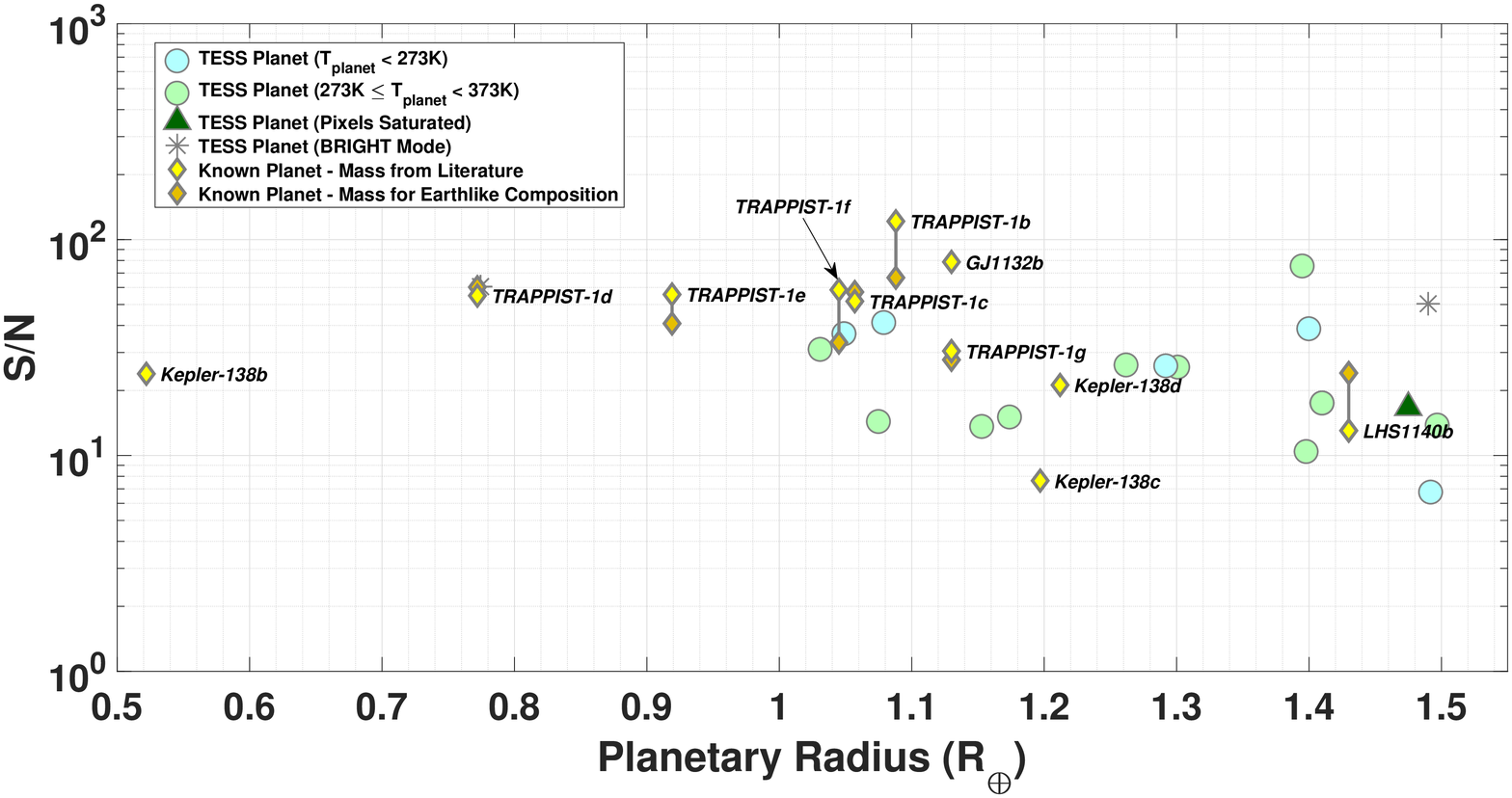}{1.0\textwidth}{}}
\caption{Integrated Signal-to-Noise (S/N) we predict for NIRISS detection of the atmosphere in 100-hour observation programs of potentially habitable anticipated TESS discoveries--planets with $R_{\rm pl} \leq 1.5R_{\oplus}$ and cool or temperate equilibrium temperatures.  We compare resulting S/N to that of known discoveries from space-based and ground-based surveys.  \label{fig:sulldatacoldtemp}}
\end{figure*}

We present our predicted NIRISS S/N for the anticipated TESS discoveries published by \cite{2015ApJ...809...77S} in Figure \ref{fig:sulldata}, overplotting our predictions of S/N for existing exoplanet discoveries from space-based and ground-based surveys.  We then present predicted S/N for more thorough 100-hour observation programs of potentially habitable TESS discoveries in Figure \ref{fig:sulldatacoldtemp}.  Our results show that TESS is likely to discover many super-Earths and sub-Neptunes ($1.5R_\oplus < R_{\rm pl} \leq 4R_\oplus$) that are more amenable to atmospheric characterization than anything we have yet discovered.  However, our results also show that for small exoplanets ($R_{\rm pl} \leq 1.5R_\oplus$) we expect very few TESS discoveries will be better for atmospheric characterization than already-discovered exoplanets.  We emphasize that this outcome is based upon Kepler-derived occurrence rates, and that co-planar compact multi-planet systems (e.g., TRAPPIST-1) may be under-represented in the predicted TESS planet yield. 

Closer examination of the small anticipated TESS planets with the highest anticipated S/N reveals the properties of the TESS discoveries likely to be most conducive to follow-on atmospheric characterization studies.  Two small planets achieved NIRISS S/N higher than TRAPPIST-1b without saturating any of the NIRISS pixels, and three cold or temperate planets with radii $\sim$$1.4R_{\oplus}$ have a higher S/N than LHS1140b.  Table \ref{tab:highSNRsull} shows that these planetary systems with high S/N values have system parameters comparable to those of GJ1132b, LHS1140b, and the TRAPPIST-1 planets, so it is not surprising that NIRISS is able to attain a high S/N.  The planets orbit nearby low radii ultra-cool or M dwarf host stars, which appear relatively bright when observed in the J-band.  In addition, from Table \ref{tab:highSNRsull} we see that the estimated radial velocity semi-amplitude for these planets is $\sim$3 m s$^{-1}$ or greater for all except the smallest planet.  Thus, we anticipate that current or forthcoming radial velocity instruments should be able to estimate the masses of these promising TESS discoveries. 

\begin{deluxetable*}{cDDDDcDcDD}
\tablecaption{Terrestrial-Sized Anticipated TESS Planets with High NIRISS S/N\label{tab:highSNRsull}}
\tablecolumns{10}
\tablewidth{0pt}
\tablehead{
\colhead{TESS Catalog} & \multicolumn2c{J} & \multicolumn2c{V} & \multicolumn2c{Distance} & \multicolumn2c{Stellar} &
\colhead{Stellar} & \multicolumn2c{Planet} & 
\colhead{Planet} & \multicolumn2c{Radial Velocity} & \multicolumn2c{NIRISS} \\ 
\colhead{Number} & \multicolumn2c{} & \multicolumn2c{} & \multicolumn2c{} & \multicolumn2c{Radius, $R_*$} &
\colhead{Temperature} & \multicolumn2c{Radius, $R_{\rm pl}$} & \colhead{Temperature} & \multicolumn2c{Semi-Amplitude} & \multicolumn2c{S/N} \\
\colhead{} & \multicolumn2c{(mag)} & \multicolumn2c{(mag)} & \multicolumn2c{(pc)} & \multicolumn2c{($R_{\odot}$)} &
\colhead{($ {\rm K}$)} & \multicolumn2c{($R_{\oplus}$)} &
\colhead{($ {\rm K}$)} & \multicolumn2c{(${\rm m \ s^{-1}}$)} & \multicolumn2c{} \\
}
\decimals
\startdata
922 & 11.3 & 16.2 & 13.2 & 0.116 & 2730 & 1.39 & 290 &6.65 & 23.8 \\
1308 & 7.97 & 11.6 & 8.71 & 0.249 & 3370 & 1.49 & 289 & 2.90 & 16.0 \\
1622 & 11.4 & 15.3 & 26.9 & 0.172 & 3230 & 0.71 & 693 & 0.76 & 30.6 \\
1745 & 10.4 & 14.0 & 25.1 & 0.224 & 3340 & 1.09 & 904 & 2.97 & 23.6 \\
1919 & 12.2 & 16.8 & 22.4 & 0.119 & 2840 & 1.40 & 236 & 5.08 & 12.2 \\
\enddata
\tablecomments{Planetary systems parameters, including the estimate of radial velocity semi-amplitude, are taken from \cite{2015ApJ...809...77S}. NIRISS signal-to-noise is calculated in this work.}
\end{deluxetable*}

Two prominent features in Figure \ref{fig:sulldata} lead us to examine the following questions.  First, why are there so many cold and temperate super-Earth and sub-Neptune planets with S/N comparable to or better than that for hot planets with similar planetary radii?  Does this make sense, given the fact that we would expect the scale height of the planetary atmosphere, and thus the signal, to increase with an increase in the temperature of the planet's atmosphere?  Second, the S/N appears relatively flat in the radius regime $1.5R_{\oplus} < R_{\rm pl} \leq 4R_{\oplus}$.  But according to equation \ref{eq:S}, the signal produced by the atmosphere should increase as density decreases, and Figure \ref{fig:comparechenkipmass} shows that density decreases with planetary radius across this regime.  Why doesn't the S/N have a positive slope on this plot?

\begin{figure*}
\gridline{\fig{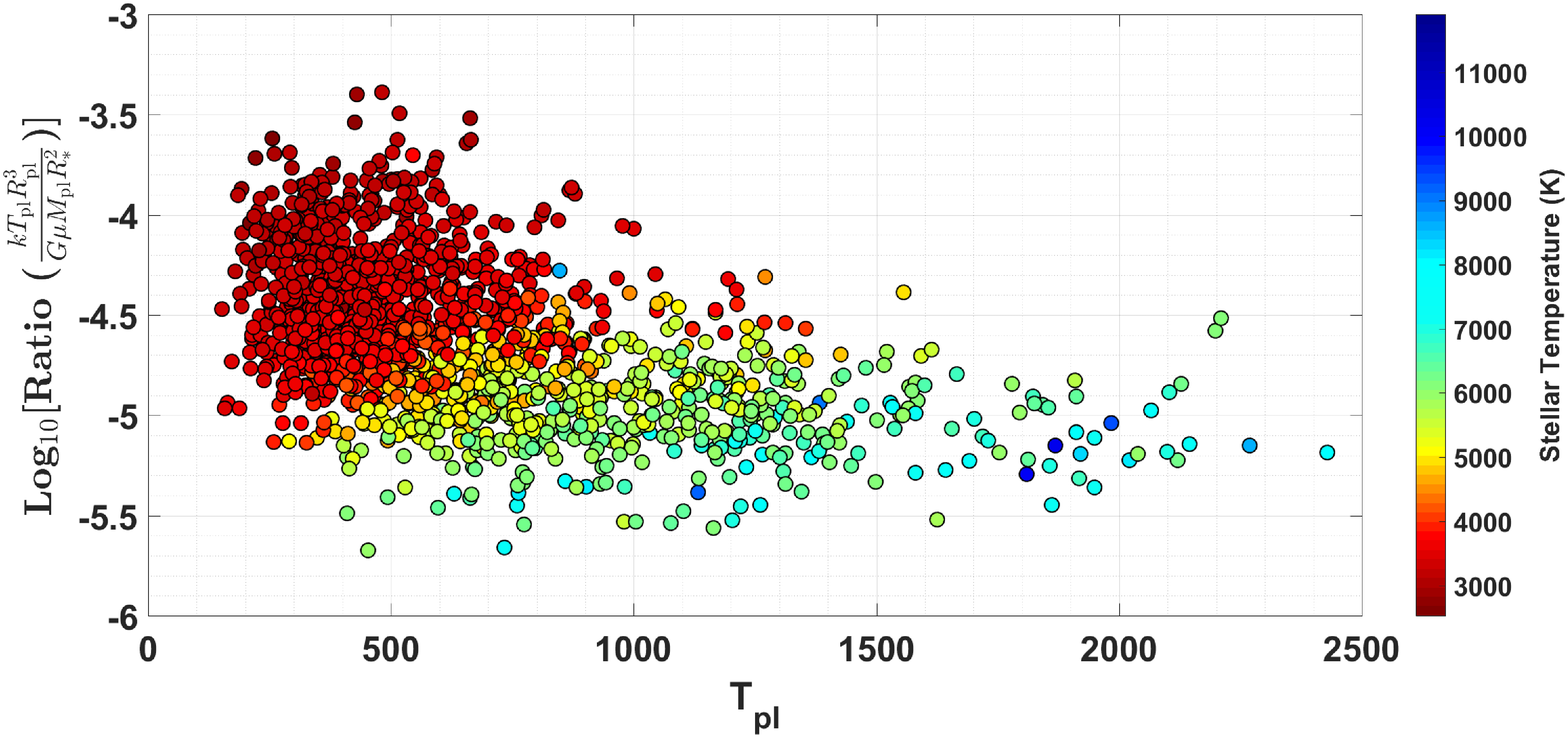}{1.0\textwidth}{}}
\caption{The signal produced by the planetary atmosphere varies directly with the planet's temperature, and inversely with the planetary density and stellar cross-sectional area.  We create a dimensionless ratio of these factors by multiplying by $\frac{k}{\mu G}$, where k is Boltzmann's constant, G is Newton's gravitational constant, and $\mu$ is the molecular weight in kg of the hydrogen-dominated atmosphere.  We then plot this dimensionless ratio versus planetary temperature for those planets with $1.5R_{\oplus} < R_{\rm pl} \leq 4R_{\oplus}$.  We see that many of the cold and temperate planets predicted by \cite{2015ApJ...809...77S} have system parameters that produce a high value of the dimensionless ratio, which accounts for the considerable number of cold and temperate planets with relatively high S/N values in Figure \ref{fig:sulldata}.  In particular, the highest values of the dimensionless ratio are produced by those planets orbiting the coolest host stars, which also have the smallest cross-sectional areas.  \label{fig:SNvsRatio}}
\end{figure*}

The considerable number of cold and temperate super-Earths and sub-Neptunes with relatively high S/N is explained by realizing that the signal depends not only on the temperature of the planet's atmosphere, but also on the planet's density and the cross sectional area of the star, as can be seen by examining equations \ref{eq:H} and \ref{eq:S}.  In Figure \ref{fig:SNvsRatio}, we plot a dimensionless ratio of planet temperature, density, and stellar cross-sectional area (the ``signal") versus planet temperature.  The colorbar on the plot indicates stellar temperature.  We see that many of the cold and temperate planets predicted by \cite{2015ApJ...809...77S} have system parameters that produce a high value of the dimensionless ratio, which accounts for the considerable number of cold and temperate planets with relatively high S/N values in Figure \ref{fig:sulldata}.  In particular, the highest values of the dimensionless ratio are produced by those planets orbiting the coolest host stars, which also have the smallest cross-sectional areas.  

\begin{figure*}
\gridline{\fig{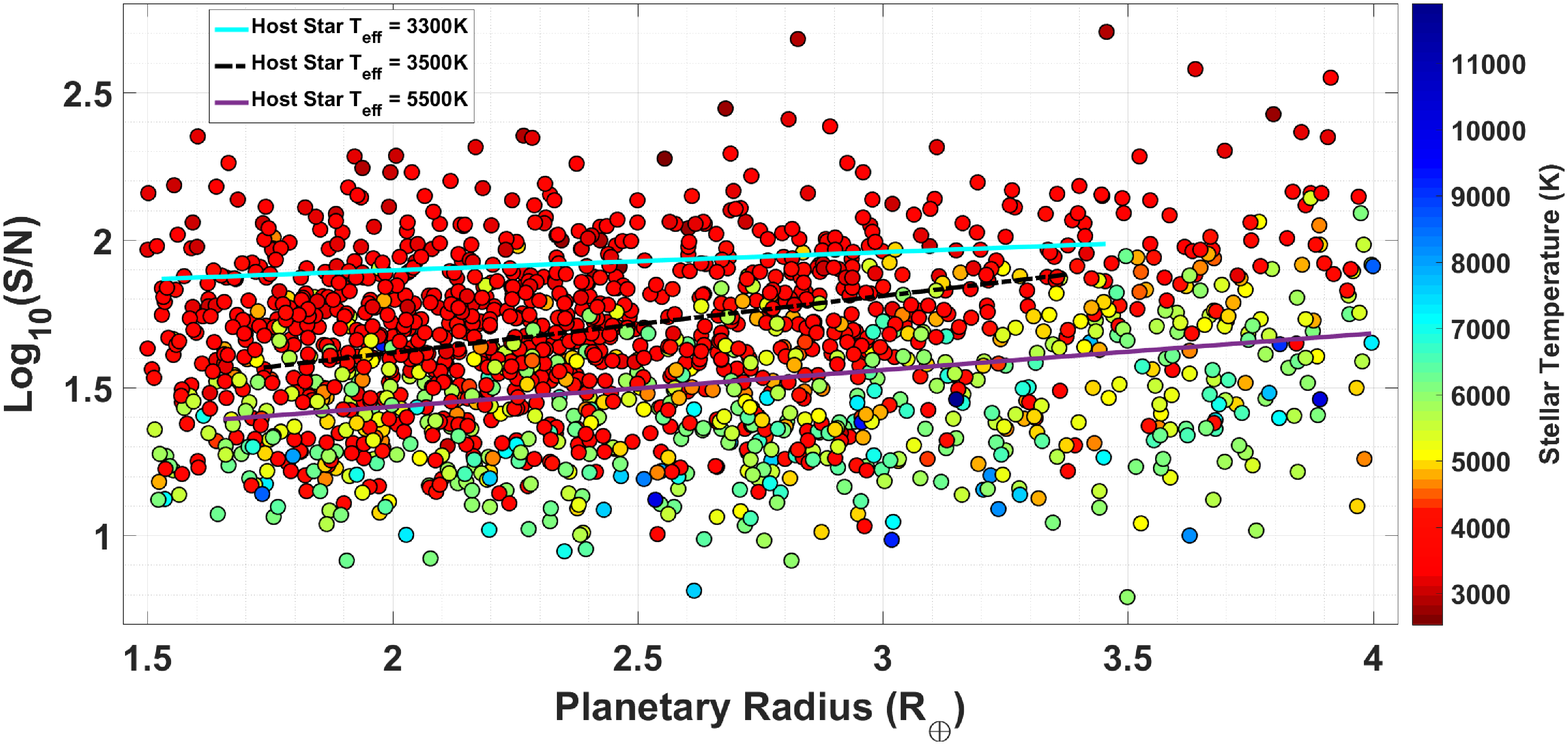}{1.0\textwidth}{}}
\caption{For a given host star, the signal produced by the planetary atmosphere during transit increases with planetary radius.  However, this increase in S/N with planetary radius is hidden in Figure \ref{fig:sulldata} since all of the TESS systems are shown on a single Figure.  In addition, the stellar types observed vary with planetary radius due to a combination of the preselected target stars chosen for the TESS mission and the planet occurrence rates employed for those target stars in the \cite{2015ApJ...809...77S} simulations.  Here, we plot S/N versus planetary radius with the color of the symbols indicating stellar temperature.  By referring to each color (i.e., each stellar type) in this Figure separately, we detect a trend towards a positive slope across the sub-Neptune radius regime.  To aid in visualization, we also plot best-fit lines for host stars of three different stellar temperatures. From top to bottom, the three lines are for host star temperatures of 3300K, 3500K, and 5500K.  Each line spans planetary radii values between the lowest and highest planetary radii values used in fitting the line.  \label{fig:SNvsRpColorbar}}
\end{figure*}

The apparent flatness of the S/N in the radius regime $1.5R_{\oplus} < R_{\rm pl} \leq 4R_{\oplus}$ of Figure \ref{fig:sulldata} results from a combination of the type of preselected target stars chosen for the TESS survey and the planet occurrence rates employed for those target stars in the \cite{2015ApJ...809...77S} simulations.  For multiple planets orbiting the same host star, we expect S/N to increase with planetary radius.  However, the TESS planets of various radii orbit different host stars, and other factors will also influence S/N.  Nevertheless, we would expect that if we plot S/N versus $R_{pl}$ for a large sample of TESS host stars of the same stellar type, the best-fit line should have a positive slope. We illustrate this in Figure \ref{fig:SNvsRpColorbar} by showing S/N versus planetary radius, with the symbols color-coded by stellar temperature.  By referring to each color (i.e., each stellar type) separately, we detect a trend towards a positive slope across the sub-Neptune radius regime. To aid in visualization, we also plot best-fit lines for host stars of three different stellar temperatures.  However, we emphasize that we are only looking qualitatively for a positive trend.  Although the host stars have the same stellar temperatures, the planetary systems vary in stellar radii, distance from Earth (affecting brightness and thus on-sky efficiency $\eta$), and planetary equilibrium temperature, all of which will influence the attainable S/N.  For the best-fit lines, we used 18 planetary systems to compute the best-fit line for 3300K, 22 planetary systems to compute the best-fit line for 3500K, and 25 planetary systems (with $5470{\rm K} \leq T_{\rm eff} \leq 5530{\rm K}$) to compute the best-fit line for 5500K.  For each of the three stellar temperatures examined, we use the same PHOENIX stellar model spectra for all host stars.

\subsection{Sensitivity to Planetary Composition} \label{subsec:resultsplanetcomp}

\begin{figure*}
\gridline{\fig{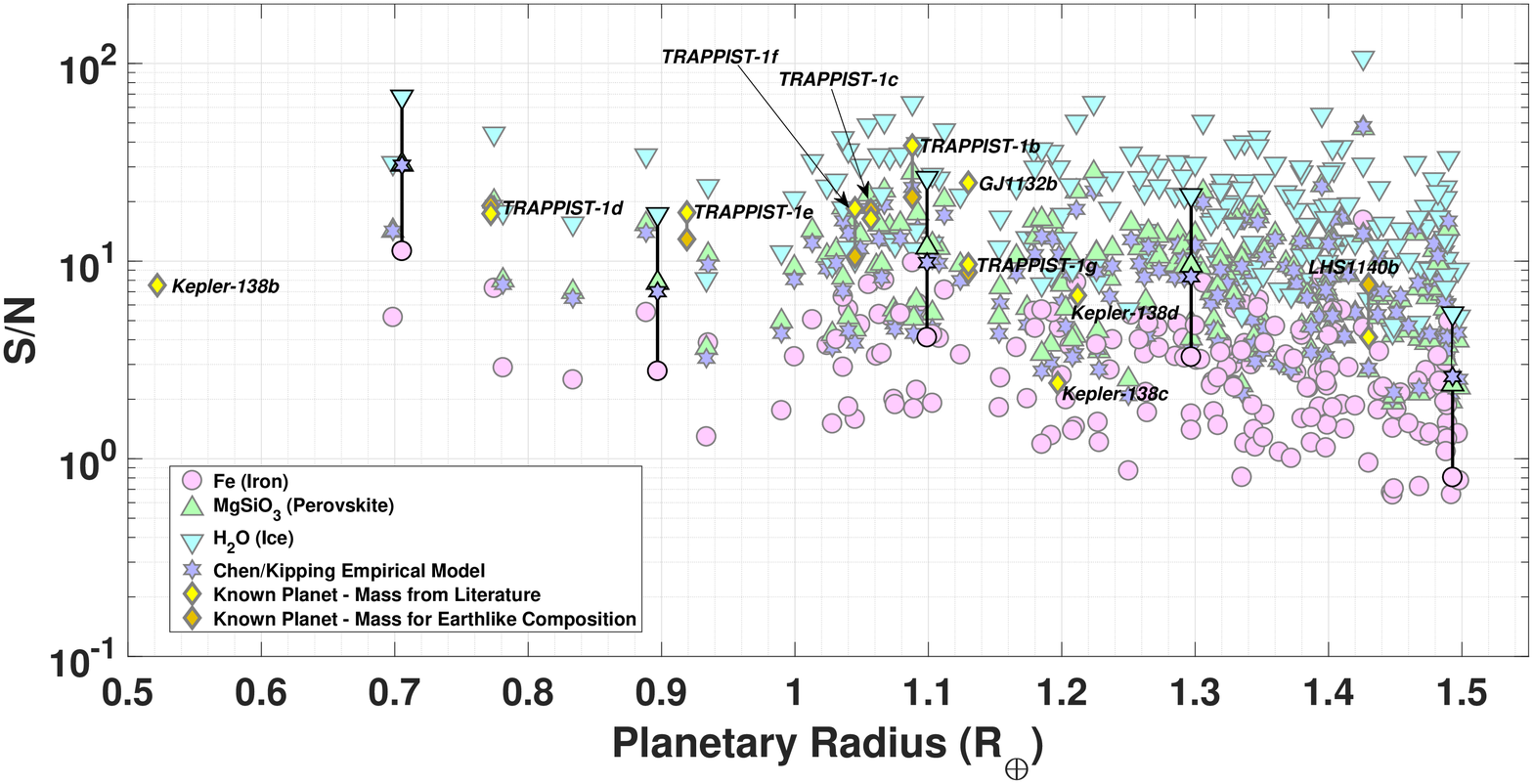}{1.0\textwidth}{}}
\caption{Integrated Signal-to-Noise (S/N) we predict for NIRISS detection of the atmosphere in 10-hour observation programs of anticipated TESS planets for four different compositions.  We calculated masses for the three homogeneously composed planets using the theoretical relationships of \cite{2007ApJ...669.1279S}, and we compared this to the S/N attainable with masses calculated using the \cite{2017ApJ...834...17C} empirical relationship.  We overplot our predictions of S/N for known exoplanets, where we used the masses reported in Tables \ref{tab:spacedisc}, \ref{tab:ground}, and \ref{tab:Earthlike}.  For the \cite{2015ApJ...809...77S} anticipated TESS planets, our predicted NIRISS S/N for an ice composition is 6 to 7 times higher than that for a dense iron composition.  To better visualize the impact of various compositions on a single planet, we chose five anticipated TESS planets and outlined the four symbols (one for each composition) in black, joining the symbols with a black line.\label{fig:sullcompSNR}}
\end{figure*}

We turn now to an examination of the sensitivity of our results shown in Figure \ref{fig:sulldata} to various factors, beginning with planetary composition.  For each TESS planet, Figure \ref{fig:sullcompSNR} shows the S/N attained for three homogeneous compositions, where we calculated masses using theoretical models \citep{2007ApJ...669.1279S}, and we compare to the S/N predicted for the \cite{2017ApJ...834...17C} empirical mass model.  For comparison, we overplot our predictions of S/N attainable for currently known exoplanets. 

In the radius regime examined, the NIRISS S/N attained for a given TESS planet with an ice composition is 6 to 7 times higher than that attained for an iron composition.  The S/N values for the pervoskite composition and Chen/Kipping mass model lie between the values for ice and iron.  The wide variation of S/N for TESS planets of different compositions emphasizes the importance of constraining planetary masses prior to conducting JWST observations \citep{2017ApJ...836L...5B}.

Low density is one factor contributing to a relatively high S/N.  Referring to Figure \ref{fig:comparechenkipmass}, we see that the relatively low estimated densities of the TRAPPIST-1 planets contribute to their high S/N.  In fact, the S/N for most of the TRAPPIST-1 planets falls when density is estimated using an Earth-like composition.  Interestingly, although the current density estimate of LHS-1142b lies near that of a dense, pure iron planet, its S/N rivals that of similarly-sized TESS planets with masses calculated using the Chen/Kipping empirical relationship.  In fact, the S/N for LHS-1142b rises when we estimate its mass using an Earth-like composition.  Thus, as the masses of these planets are further constrained, we would expect our predictions of S/N to change. 

\subsection{Observational Overhead Effects} \label{subsec:resultsobsoh}

\begin{figure*}
\gridline{\fig{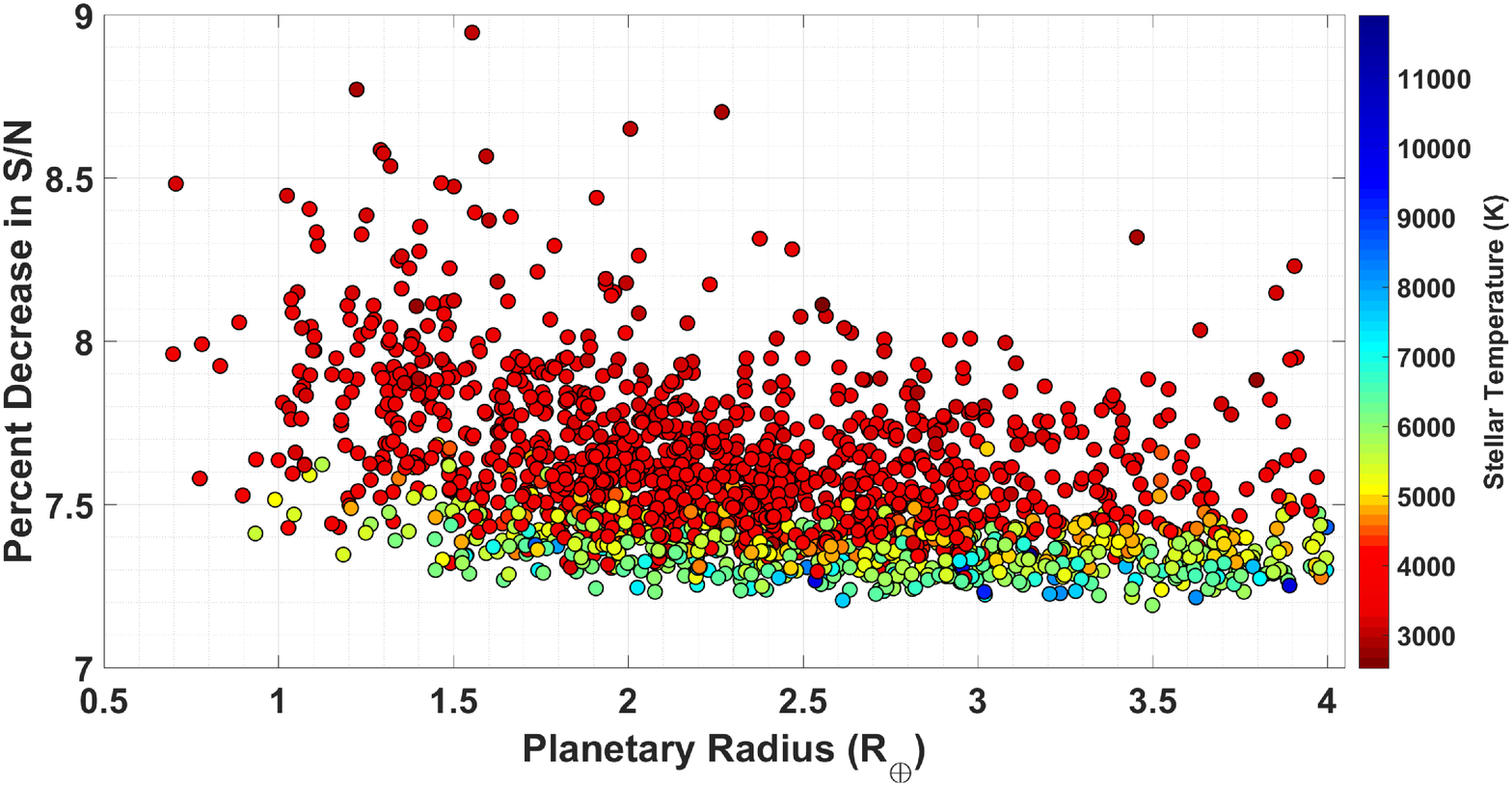}{1.0\textwidth}{}}
\caption{Our predictions of the percent decrease in Signal-to-Noise (S/N) when considering observational overhead for 10-hour observation programs.  In general, the coolest stars suffer the highest percent decrease in S/N.  This is because the transit duration is generally shorter for cooler--and thus smaller--stars, where the orbital semi-major axis is shorter.  The decrease in the number of cool host stars with planetary radius results from a combination of a selection bias for M dwarfs in the preselected TESS target stars, the planet occurrence rates used in the \cite{2015ApJ...809...77S} simulations, and increased TESS sensitivity to shorter period exoplanets. \label{fig:sullobsOH}}
\end{figure*}

The S/N estimates presented thus far assume 10-hour observation programs, with equal amounts of time in and out of transit.  Figure \ref{fig:sullobsOH} shows that when observational overhead is considered in 10-hour observation programs, we will have a 7 to 9 percent decrease in predicted S/N for the TESS anticipated discoveries, with the coolest host stars suffering the largest impacts.  Planetary systems with short transit durations suffer greater effects from the \textit{constant time periods per visit} required to set up the JWST observatory or the NIRISS instrument.  The reason for this impact is that planetary systems with short transit durations must be revisited more often in order to accumulate a given amount of scientific observation time.  Thus, a greater amount of the total requested telescope time is devoted to non-scientific activities.  Note that the apparent decrease in the number of cool host stars with planetary radius in Figure \ref{fig:sullobsOH} is due to a combination of a selection bias for M dwarfs in the preselected TESS target stars, the planet occurrence rates used in the \cite{2015ApJ...809...77S} simulations, and increased TESS sensitivity to shorter period exoplanets.   

\subsection{Cloud Effects} \label{subsec:resultsclouds}

\begin{figure*}
\gridline{\fig{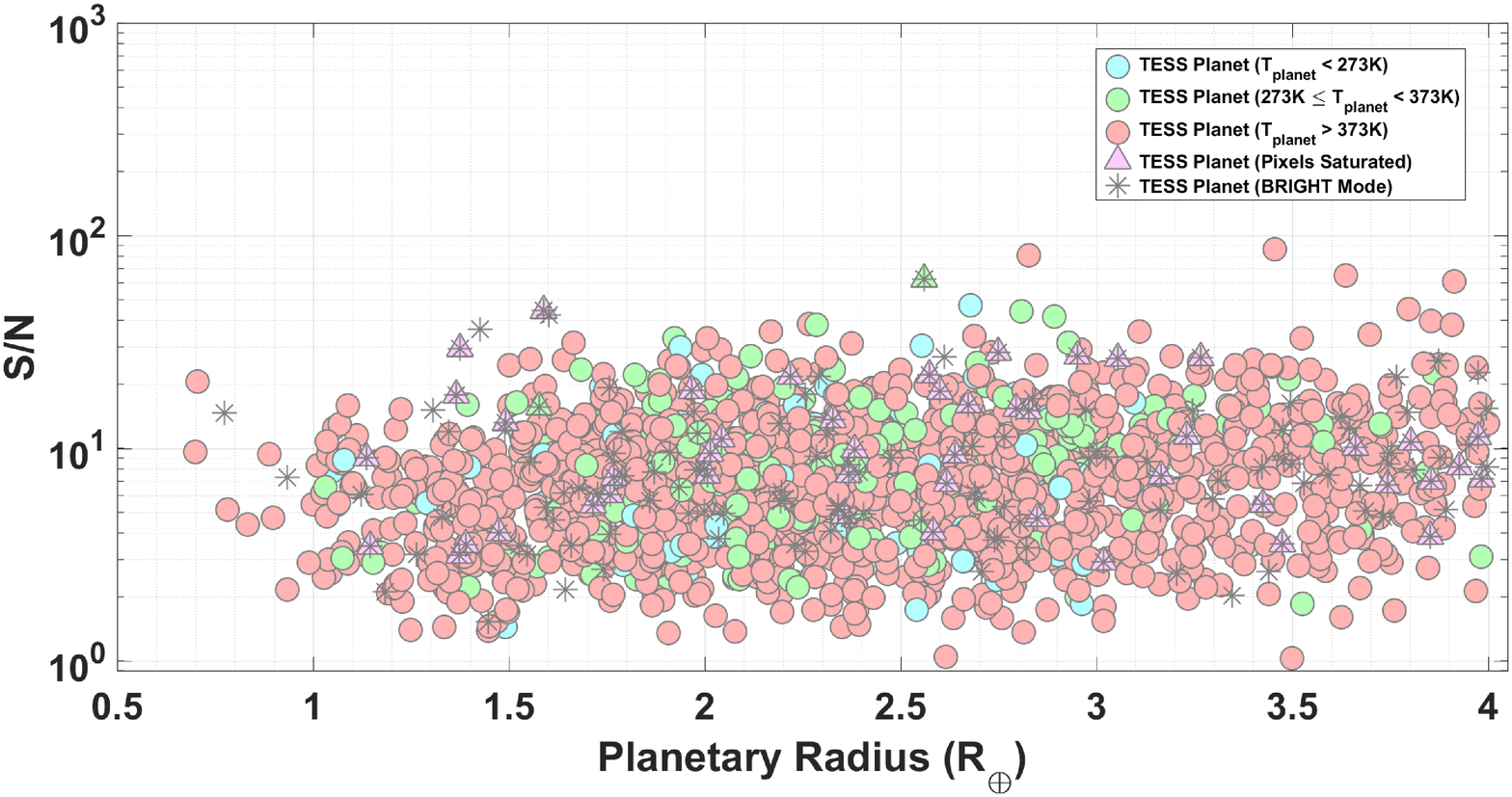}{1.0\textwidth}{}}
\caption{Our predictions showing the anticipated Signal-to-Noise (S/N) achievable in 10-hour observation programs of planets with cloud decks placed at a pressure of 10 mbar.  When compared to our results in Figure \ref{fig:sulldata}, cloudy atmospheres will reduce S/N values for small planets with water atmospheres ($R_{\rm pl} \leq 1.5R_\oplus$) by about 1.5 times, and for larger planets with hydrogen-dominated atmospheres ($1.5 < R_{\rm pl} \leq 4R_\oplus$) by about 5.5 times.  \label{fig:sullcloud}}
\end{figure*}

The presence of clouds in planetary atmospheres decreases S/N by effectively blocking transmission of stellar radiation above some pressure level, allowing us to detect spectral features due to molecular absorption only in regions of the atmosphere above the level where clouds condense.  As shown in Figure \ref{fig:sullcloud}, we estimate the effects of clouds on S/N by placing an opaque cloud deck at 10 mbar.  For small planets with water atmospheres ($R_{\rm pl} \leq 1.5R_\oplus$), the S/N for  clear atmospheres shown in Figure \ref{fig:sulldata} is about 1.5 times higher than that found for cloudy atmospheres.  For larger planets with hydrogen-dominated atmospheres ($1.5R_\oplus < R_{\rm pl} \leq 4R_\oplus$), we find that S/N for clear atmospheres is about 5.5 times greater than that for cloudy atmospheres.  

During actual observations, the percentage of cloud cover as well as the pressure level where clouds condense will vary depending upon the observed exoplanet.  In Section \ref{sec:JWSTfollowup}, we explore this statistical variation in cloud effects by applying random fluctuations to the number of atmospheric scale heights through which water absorption can be detected.

\subsection{Analysis Of 50 Trials of TESS Primary Mission} \label{subsec:results50trials}

\begin{figure*}
\gridline{\fig{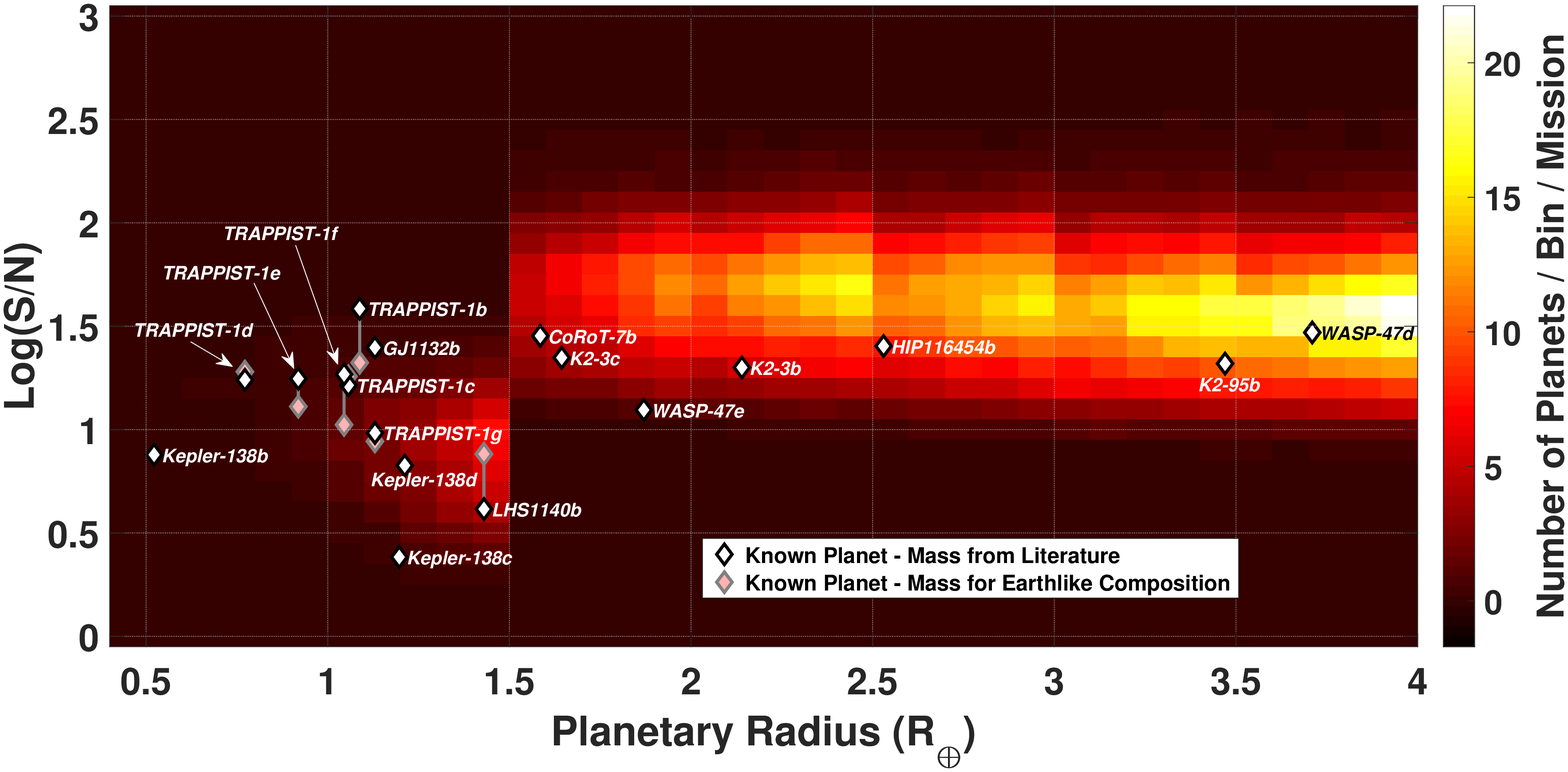}{1.0\textwidth}{}}
\caption{Two-dimensional histogram showing the Signal-to-Noise (S/N) we predict for 10-hour observation programs based upon analysis of 50 Monte Carlo realizations of the TESS primary mission.  Thirty bins are logarithmically spaced from $10^0$ to $10^3$ along the S/N axis, and bins are spaced every $0.1R_\oplus$ along the $R_{\rm pl}$ axis.  For comparison, we overplot our results for existing discoveries from space-based and ground-based surveys.\label{fig:50trial}}
\end{figure*}

The catalog of anticipated TESS discoveries published by \cite{2015ApJ...809...77S} represents only one possible outcome for the planet yield from the TESS primary mission.  To achieve a more statistically robust prediction of the suitability of TESS primary mission discoveries to atmospheric characterization, we examined data from 50 Monte Carlo realizations of the TESS primary mission. We present our results as Figure \ref{fig:50trial}, a 2-dimensional histogram in S/N-$R_{\rm pl}$ space. We place thirty bins logarithmically from $10^0$ to $10^3$ along the S/N axis, and every $0.1R_\oplus$ along the $R_{\rm pl}$ axis.  The number of planets found in a given bin is an average of the number predicted over all 50 trials of the TESS primary mission, and represents the expected number of planets we will discover in that regime.  As in Figure \ref{fig:sulldata}, we overplot our predictions of S/N for existing discoveries from space-based and ground-based surveys.  Figure \ref{fig:50trial} supports our conclusions from Figure \ref{fig:sulldata}.  While TESS will discover very few terrestrial-sized planets more amenable to atmospheric characterization than those that have already been discovered, TESS \textit{is} likely to find many planets with promising properties in the radius regime $1.5R_\oplus < R_{\rm pl} \leq 4R_\oplus$.  However, we reiterate that this outcome is based upon Kepler-derived occurrence rates, and that co-planar compact multi-planet systems (e.g., TRAPPIST-1) may be under-represented in the predicted TESS planet yield.   

Numerical integration of various regions of Figure \ref{fig:50trial} provides some quantitative insight into our conclusions.  For example, if we integrate the region with $R_{\rm pl} < R_{\rm GJ1132b}$ and ${\rm S/N} > {\rm S/N}_{\rm GJ1132b}$, we find that TESS is likely to discover only 1.84 planets in this regime over its 2-year primary mission.  Similarly, TESS is likely to discover only about 6.18 cold or temperate planets ($T < 373 {\rm K}$) with radii less than and S/N higher than LHS1140b with an Earthlike density.  In the radius regime $1.5R_\oplus < R_{\rm pl} \leq 2R_\oplus$, TESS will discover about 245 planets over its primary mission with S/N greater than that of K2-3c.  Similarly, in the radius regime $2R_\oplus < R_{\rm pl} \leq 2.5R_\oplus$, we predict TESS will discover about 391 planets with S/N greater than that of K2-3b.  Thus, TESS is likely to discover many promising super-Earth and sub-Neptune-sized exoplanet targets.   

\subsection{Sensitivity to Systematic Noise} \label{subsec:resultsnoisefloor}

\begin{figure*}
\gridline{\fig{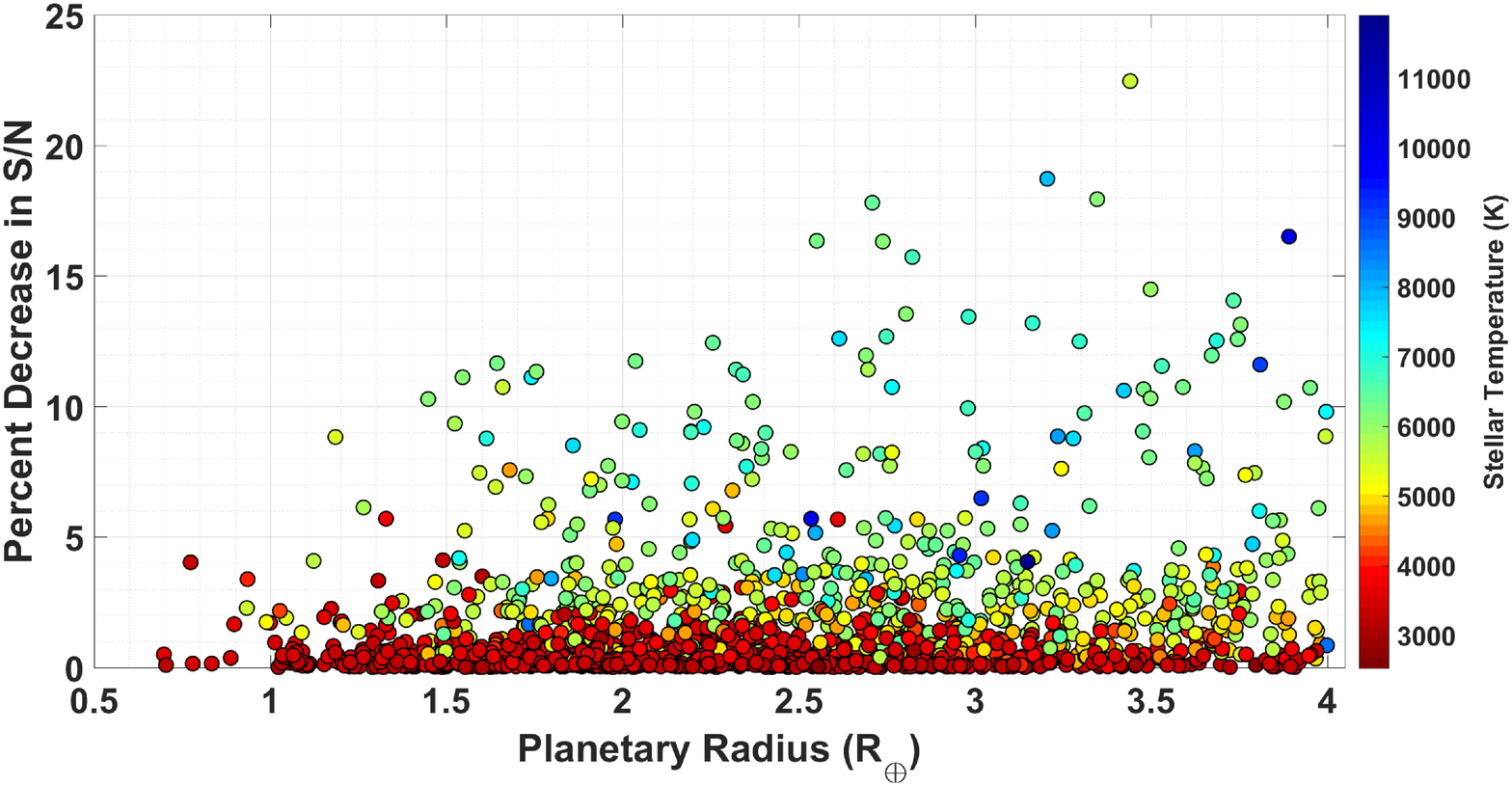}{1.0\textwidth}{}}
\caption{Our predictions of the percent decrease in Signal-to-Noise (S/N) when a 20 ppm residual noise level--decreasing with the square root of the number of transits--is incorporated into 10-hour observation programs.  The decrease in S/N is not uniform across all \cite{2015ApJ...809...77S} anticipated TESS discoveries, but varies with planetary system properties, reaching up to $\sim$20\% for some systems.  In 10-hour observation programs, the hottest host stars are affected the most by systematic noise.  This is partially because the orbital period and transit duration of TESS discoveries are longest for the hottest host stars, so that fewer visits are required to complete a given observation program.  In addition, the brightest host stars suffer the greatest effects. \label{fig:sullnfperdec}}
\end{figure*}

We present our estimates of NIRISS S/N sensitivity to systematic noise for the anticipated TESS discoveries in Figure  \ref{fig:sullnfperdec}.  For 10-hour observation programs, we find that systematic noise will cause S/N to decrease by up to $\sim$20\% (i.e., S/N is $\sim$0.8 times that without systematic noise), with the hottest host stars suffering the greatest effects.  One reason that the hottest stars are affected most is that the TESS discoveries with the hottest host stars also generally have the longest orbital periods and the longest transit durations, so that fewer visits are required to accumulate 10 hours of observation time.  Since the systematic noise decreases with the square root of the number of observed transits, the hottest host stars are thus affected more.  Brighter host stars also generally suffer greater effects.  

\begin{figure*}
\gridline{\fig{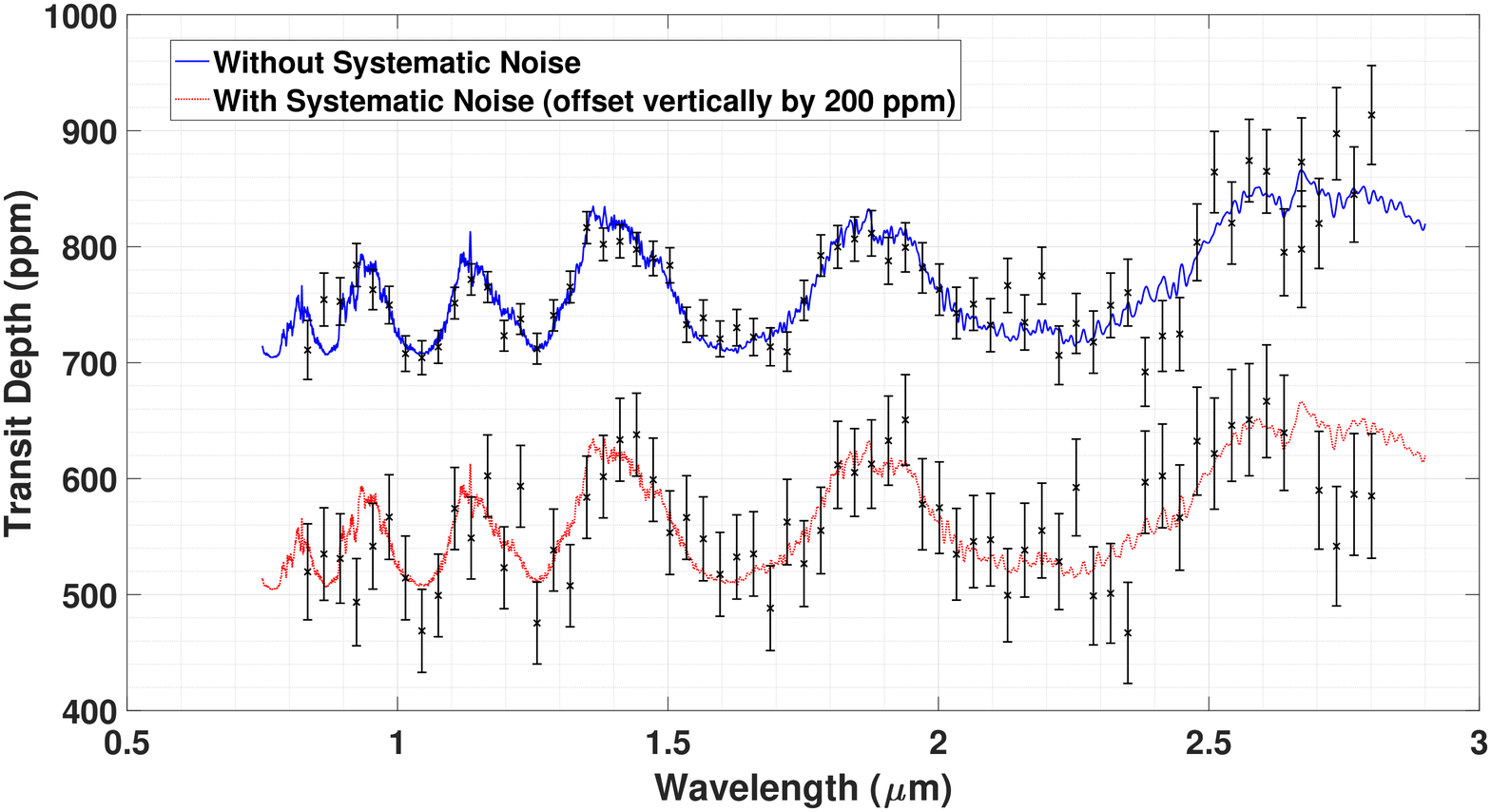}{1.0\textwidth}{}}
\caption{Predicted transit spectra resulting from a 10-hour observation program of the known exoplanet K2-3c, assuming a hydrogen-dominated atmosphere. Spectra for the planet are shown both with and without systematic noise.  The systematic noise is set at 20 ppm for one transit, but decreases as the square root of the number of transits.  Although systematic noise will affect the quality of the observed spectrum, the variation of transit depth with wavelength should be detectable.   \label{fig:specPA}}
\end{figure*}

Importantly, we note that the spectral scale at which we apply the systematic noise is critical.  In our results for the 1,984 \cite{2015ApJ...809...77S} TESS planets, we applied the systematic noise to the resolution elements (i.e., two columns) across the NIRISS bandpass.  If instead the systematic noise is applied to larger bins (e.g., 32 columns) across the NIRISS bandpass, the effects of the systematic noise are magnified.  In Figure \ref{fig:specPA}, we show the results of applying systematic noise to the known exoplanet K2-3c.  As described in Section \ref{subsubsec:noisefloor}, we apply the systematic noise to each of the 64 bins in our spectra.  For K2-3c, we also tried applying the systematic noise to each NIRISS resolution element (i.e., two columns) instead, and we found that in that case the effects of the systematic noise on each of the 64 bins was negligible.

We use estimated S/N for a 10-hour observation program of K2-3c, which we assume has a hydrogen-dominated atmosphere, to develop the predicted transmission spectra of Figure \ref{fig:specPA}.  In the S/N presented in previous sections, photon noise constitutes the major noise source.  In our predicted spectral plot for K2-3c, we show two spectra: one where photon noise dominates, and one with systematic noise incorporated in addition to all other sources of noise.  Although the addition of systematic noise does affect the output spectra in Figure  \ref{fig:specPA}, the variation of transit depth with wavelength--as caused by water lines in an atmosphere--will still be detectable in a 10-hour observation program.  

\section{JWST Follow-Up Observation Program} \label{sec:JWSTfollowup}

We have shown in Sections \ref{subsec:resultsSNRsullivan} and \ref{subsec:results50trials} that TESS will discover many promising exoplanet targets for atmospheric characterization in the radius regime $1.5R_\oplus < R_{\rm pl} < 2R_\oplus$.  Recent work has shown that planetary atmospheres are likely to transition from hydrogen-dominated atmospheres to high molecular weight atmospheres within this radius regime \citep{2015ApJ...801...41R, 2017arXiv170310375F}.  Here, we use our results from Section \ref{sec:results} to estimate the scope of a JWST follow-up observation program of TESS discoveries that would enable us to map the transition from low to high mean molecular weight atmospheres.  

We have run 100,000 trials of a synthetic observing program that seeks to map the nature of the transition from low to high molecular weight atmospheres.  For each trial, we compare our synthetic observations to two possible functions that describe the transition in water vapor absorption going from exo-Neptunes to the domain of rocky planets at small radii.  Since our simulations use a step function in atmospheric composition with the discontinuity at 1.5$R_\oplus$, that step function is our first possible transition function.  We compare it to a function wherein the water absorption measured in equivalent scale heights (see below) is assumed to be a linear function of the log of planetary radius, similar to the power law described by \cite{2014Natur.505...69K}.  For each trial observing program, we calculate the Bayesian Information Criterion for the step function, and the power law after fitting to the data.  Since our synthetic data are based on the step function, a sufficiently intense JWST transit spectroscopy program should produce a BIC value exceeding (by $> 10$) the BIC value for the best fitting power law.

\begin{figure}[h!]
\gridline{\fig{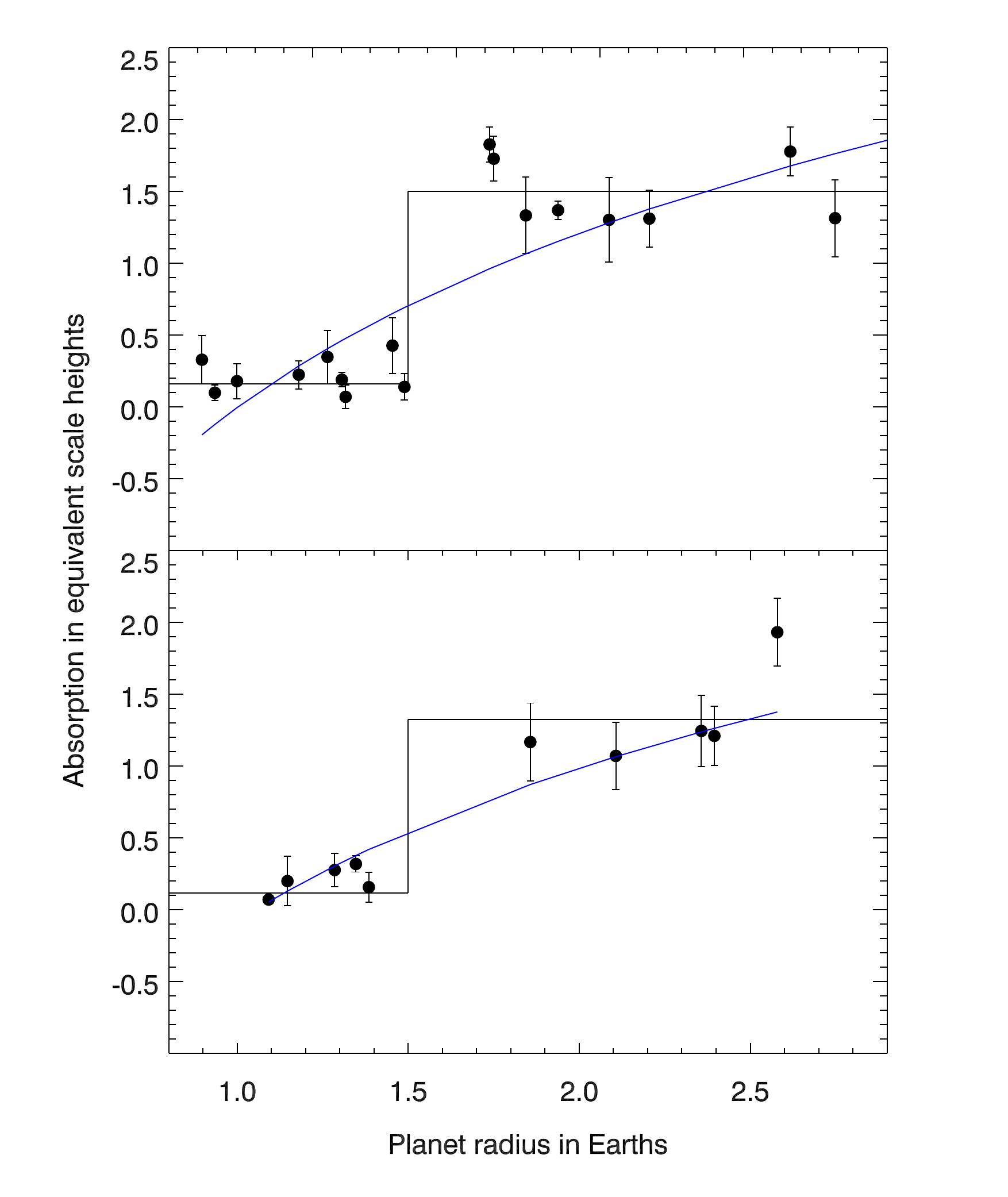}{0.5\textwidth}{}}
\caption{Examples of results from two of our trial synthetic JWST observing programs using NIRISS.  The abscissa is planet radius, and the ordinate is the equivalent number of scale heights of absorption, scaled to a mean molecular weight of 2.3.  Since the small planets have water vapor atmospheres, their equivalent scale height values are small, reflecting the high molecular weight of their atmospheres. The top panel observes 18 planets in 117 hours of charged JWST time, at one transit per planet.  It easily discriminates the step function from the poorly-fitting power law (blue line), with a BIC difference of 180.  The lower panel shows results from a 66-hour observing program, wherein the power law (blue line) is a better fit (BIC smaller by 15) than the step function, in spite of the fact that these synthetic data were drawn from a distribution using the step function.
\label{fig:Drake1}}
\end{figure}

Our trials seek to incorporate realistic observing conditions as much as possible.  The primary limitation will be due to the presence of clouds on the exoplanets.  For hot Jupiters, clouds reduce the magnitude of the water absorption from $\sim$\,5 scale heights in a clear atmosphere to much less.  In their statistical study, \citet{2016ApJ...823..109I} found an average of 1.8 scale heights of absorption in hot Jupiters, and \citet{2017ApJ...847L..22F} found 1.4 scale heights with a larger sample.  We therefore scale our calculated spectra and S/N ratios to the equivalent of $N$ scale heights for each planet, and we vary $N$ by adding random fluctuations to mimic the cosmic variation in cloud occurrence.  (This assumes that the cloud properties of small planets are statistically similar to the hot Jupiters, but inadequate statistics for small planets allow no alternative.)  We adopt a log-normal distribution for $N$, using the average value (1.4) and dispersion ($\sigma = 0.13$ in ${\rm log}_{10}$), from \cite{2017ApJ...847L..22F}. For each trial, we choose 2 to 15 planets orbiting stars brighter than J=10, picking an equal number randomly from both ranges in radii ($< 1.5R_\oplus$ and $\ge 1.5R_\oplus$). We observe one transit of each planet, we include JWST's charged overhead per visit to each transit based on Eq.~12, and we convert our scaled synthetic spectra and signal-to-noise ratios to an equivalent number of scale heights, under the assumption that every planet has a mean molecular weight of 2.3 (i.e., a H-He atmosphere).  This scaling is not physical, but conveniently serves to illustrate the transition between the H-He and water vapor atmospheres.  

We fit our transition functions to the number of equivalent scale heights of absorption as a function of planetary radius.  The number of equivalent scale heights effectively measures the composition of the atmosphere, because the high molecular weight atmospheres will produce smaller signals. Our simulated observing program interprets those small signals as fewer equivalent scale heights.  This simple method follows \citet{2009ApJ...690.1056M}, and is a conservative (worst-case) procedure because more sophisticated analysis methods (e.g., retrievals for all planets as per \citealt{2013ApJ...775..137L}) would have greater ability to clarify the nature of the transition function.

Figure~\ref{fig:Drake1} shows two example trials from our simulation.  The top panel shows a 117-hour observing program that obtains spectra of 9 planets in each radius regime (18 planets total), and it easily discriminates the step function from the power law.  However, the lower panel shows a 66-hour observing program (10 total planets) that would conclude in favor of the power law, in spite of the fact that the synthetic data are derived from the step function.  That occurs because two of the rocky planets have very small observed errors, and by chance the power law that connects them extends reasonably well to the larger planets.  Since the planets with the smallest errors dominate $\chi^2$ (and hence the BIC), the result in this case would erroneously conclude in favor of a power law relation between planet radius and atmospheric mean molecular weight.

\begin{figure}[h!]
\gridline{\fig{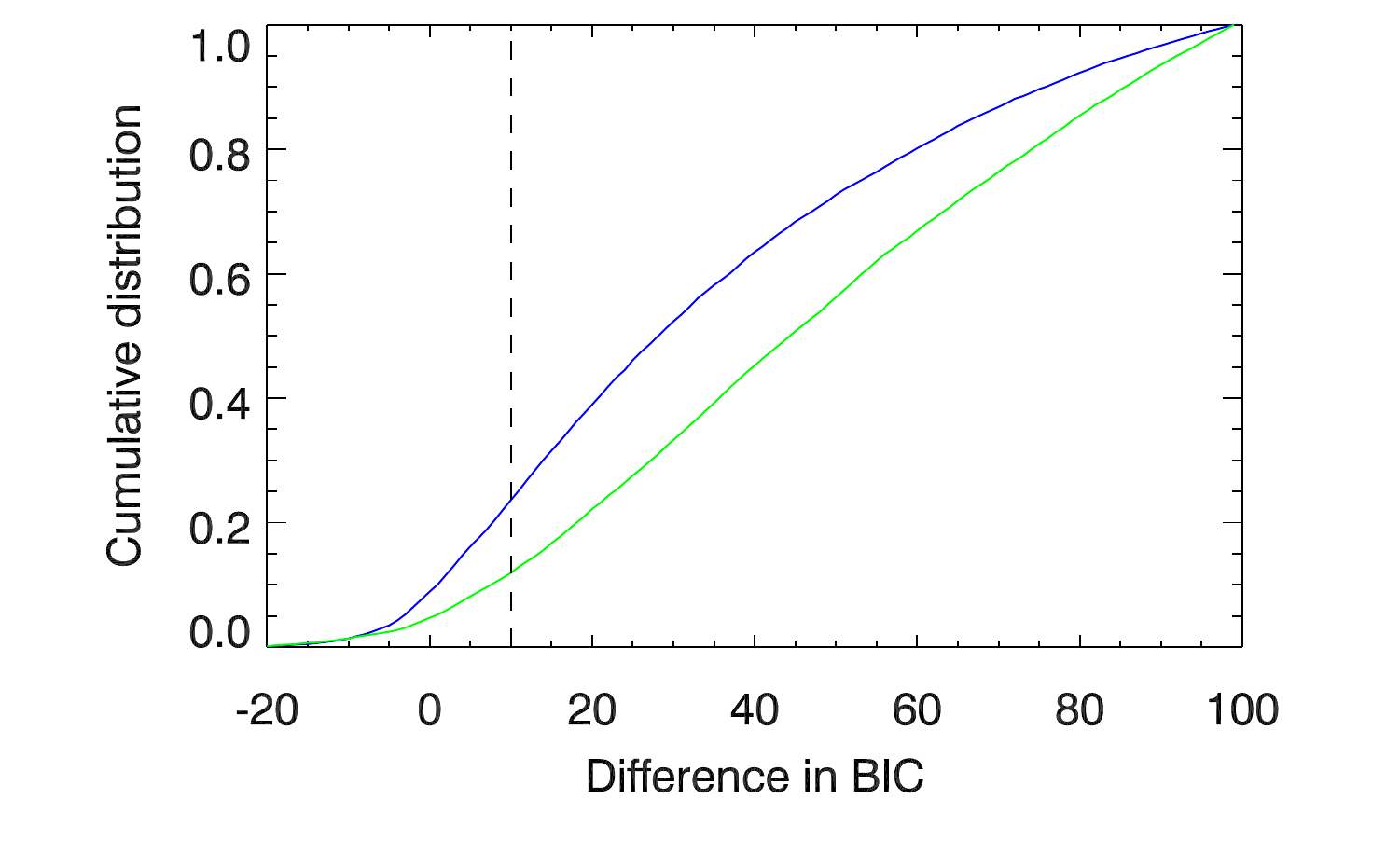}{0.5\textwidth}{}}
\caption{Cumulative distribution of the difference in Bayesian Information Criteria for best fit step functions and the power law, wherein a difference greater than 10 (vertical dashed line) strongly favors the step function.  The blue line shows the distribution for observing programs between 60 and 100 hours of charged JWST time, and the green line is for programs between 100 and 140 hours.
\label{fig:Drake2}}
\end{figure}

Considering the totality of our 100,000 trials, we find good news, and bad news, both illustrated in Figure~\ref{fig:Drake2}.  The good news is that the TESS planets will enable even modest observing programs (between 60 and 100 hours of charged time) to define the physical nature of the transition between low and high molecular weight atmospheres.  Specifically, the cumulative distribution of ${\Delta}BIC$ values for the 60-100 hour observing programs exceeds 10 (strongly supporting the step function) 76\% of the time, and 100-140 hour observing programs have ${\Delta}BIC$ exceeding 10 for 88\% of the trials.  The bad news is that even the largest programs fall short of the levels of certainty that are desired for such an important problem (we would prefer 95\% or greater).  However, inspection of the failed cases (as in the lower panel of Figure~\ref{fig:Drake1}) indicates how to achieve greater certainty.  Observing programs often fail when a relatively few planets are observed to high signal-to-noise, and they dominate the BIC values, but the random nature of cloud coverage biases the best fitting function.  Specifically, a cloudy atmosphere on a small planet combined with a clear atmosphere on a larger planet, can be mis-interpreted as a trend in mean molecular weight.  Other caveats are that the transition from low- to high-molecular-weight atmospheres may be more complex than either of our simple models, and the nature and occurrence of clouds may differ from the statistical behavior that we have inferred based on current observations.  Nevertheless, based on our simple assumptions, we conclude that good observing program design (uniformity in observed signal-to-noise from planet to planet), combined with analysis methods that break degeneracies between water abundance and cloud coverage, can potentially enable JWST observing programs of 60 to 100 hours to define the nature of the transition in atmospheric properties going from Neptunes to rocky super-Earths. 

\section{Summary and Conclusion} \label{sec:conclusion}

We have analyzed the anticipated TESS discoveries published by \cite{2015ApJ...809...77S}, as well as 50 Monte Carlo realizations of the TESS primary mission produced by \cite{2017arXiv170508891B}, to predict the NIRISS S/N likely to be achieved in transmission spectroscopy of the TESS planets.  In order to predict the TESS discoveries with the best prospects for atmospheric characterization, we compared our results to  predictions of S/N for 18 already known exoplanets.  In calculating S/N, we modeled all $R_{\rm pl} \leq 1.5R_{\oplus}$ planets with cloud-free pure water atmospheres, and all $1.5R_{\oplus} < R_{\rm pl} < 4R_{\oplus}$ planets with clear hydrogen-dominated atmospheres.  We investigated the effects of planetary composition, cloud cover, and systematic noise on our results.  We found:

\begin{enumerate}
\item TESS will find hundreds of super-Earths and sub-Neptunes ($1.5R_\oplus < R_{\rm pl} \leq 4R_\oplus$) capable of producing higher S/N than already known exoplanets.  In particular, TESS is likely to produce $\sim$245 planets within the radius regime ($1.5R_\oplus < R_{\rm pl} \leq 2R_\oplus$) with S/N higher than that of K2-3c, and $\sim$391 planets within the radius regime ($2R_\oplus < R_{\rm pl} \leq 2.5R_\oplus$) with S/N higher than that of K2-3b.
\item TESS will find very few terrestrial-sized planets ($R_{\rm pl} \leq 1.5R_\oplus$) with NIRISS S/N better than that of already-known exoplanets such as the TRAPPIST-1 planets, GJ1132b, or LHS1140b.  However, we note that the \cite{2015ApJ...809...77S} and \cite{2017arXiv170508891B} anticipated TESS discoveries are based upon the Kepler planet occurrence rates reported by \cite{2013ApJ...766...81F} and \cite{2015ApJ...807...45D}, and that co-planar compact multi-planet systems orbiting ultra-cool dwarf stars (e.g., TRAPPIST-1) may be under-represented.  Such systems may boost the number of TESS-discovered rocky planets producing high NIRISS S/N in transmission spectroscopy.
\item Our analysis of planetary composition shows that attainable S/N is sensitive to planet density.  NIRISS S/N for an ice composition is 6 to 7 times higher than that for a dense iron composition.  Thus, we emphasize the importance of constraining planet mass prior to conducting JWST follow-up observations for a given planet.
\item The presence of an opaque cloud deck at 10 mbar will decrease attainable S/N by about 5.5 times for planets with hydrogen-dominated atmospheres.  Longer observation programs will be required to constrain molecular abundances in planets with clouds.  In addition, \cite{2016ApJ...817...17G} showed that 1-11 $\micron$ spectra, requiring multiple JWST instruments, may be necessary to fully constrain cloudy atmospheres.
\item Residual systematic noise will decrease attainable S/N in 10-hour observation programs by up to $\sim$20\%, with hotter and brighter host stars suffering the most effects.  We assume the systematic noise will decrease with the square root of the number of observed transits, so effects will be minimized in programs that observe more transits within a given total duration.  We applied the systematic noise to each resolution element (i.e, 2 columns) when examining its effects on the TESS planets.

\item We applied our NIRISS S/N calculations to estimate the required magnitude of a JWST follow-up program devoted to mapping the transition region between high molecular weight and hydrogen-dominated atmospheres.  We conclude that the TESS planets will allow relatively modest observing programs (60 to 100 hours of charged JWST time) to define the nature of that transition (e.g., step function versus a power law).  However, it will be necessary to design the observing program to have good uniformity in S/N, so that the results are not biased by a few planets with high S/N and unusual atmospheric conditions (e.g., cloud coverage).
\end{enumerate}

\acknowledgments

The authors would like to thank Professor David Charbonneau and other members of the TESS Science Team for valuable discussions and inputs throughout the completion of this project.  We also thank an anonymous referee for detailed and helpful comments that improved this work.  D.L. acknowledges support from NASA Headquarters under the NASA Earth and Space Science Fellowship (NESSF) Program - Grant NNX16AP52H.  D.D. acknowledges Co-Investigator support from the TESS project.

\bibliography{JWST-NIRISSpaper}



\end{document}